\documentclass[prl,twocolumn]{revtex4}
\usepackage{amsmath}
\usepackage{epsfig}
\usepackage{float}
\usepackage{dcolumn} 
\usepackage{multirow} 
\usepackage{mathtools}
\usepackage[utf8]{inputenc}

\DeclareMathAlphabet{\mathpzc}{OT1}{pzc}{m}{it}
\newcommand{\be}{\begin{equation}}
\newcommand{\ee}{\end{equation}}
\newcommand{\bea}{\begin{eqnarray}}
\newcommand{\eea}{\end{eqnarray}}
\newcommand{\beas}{\begin{eqnarray*}}
\newcommand{\eeas}{\end{eqnarray*}}

\begin{document}
	\title
{Spectral statistics for ensembles of various real random matrices}    
\author{Sachin Kumar$^1$ and Zafar Ahmed$^2$}
\affiliation{$~^1$Theoretical Physics Section,$~^2$Nuclear Physics Division \\ Bhabha Atomic Research Centre, Mumbai 400 085, India}
\email{1: sachinv@barc.gov.in, 2:zahmed@barc.gov.in}
\date{\today}
\begin{abstract}
	We investigate spacing statistics for ensembles of various real random matrices where the matrix-elements have various Probability Distribution Function (PDF: $f(x)$) including Gaussian. For two 
modifications of $2 \times 2$ matrices with various PDFs, we derived that spacing distribution $p(s)$ of adjacent energy eigenvalues are distinct. Nevertheless, they show the linear level repulsion near $s=0$ as $\alpha s$ where $\alpha$ depends on the choice of the PDF. We also derive the distribution of eigenvalues $D(\epsilon)$ for these matrices. We construct ensembles of $1000$, $100 \times 100$ real random matrices $R$, $C$
	(cyclic) and $T$ (tridiagonal) and  real symmetric matrices: ${\cal R}'$,
	${\cal R}=R+R^t$, ${\cal Q}=RR^t$, ${\cal C}$ (cyclic), ${\cal T}$ (tridiagonal), $T'$ (pseudo-symmetric Tridiagonal), $\Theta$  (Toeplitz) , ${\cal D}=CC^t$ and ${\cal S}=TT^t$. We find that the spacing distribution of the adjacent levels of matrices
	${\cal R}$ and ${\cal R}'$ under any symmetric PDF of matrix elements is $p_{AB}(s)=A s e^{-Bs^2}$
	which approximately conforms to the Wigner surmise as $A/2 \approx B \approx \pi/4$. But under asymmetric PDFs we observe $A/2 \approx B >>\pi/4$, where $A,B$ are also sensitive to the choice of the matrix and  the PDF. More interestingly, the real symmetric matrices ${\cal C}, {\cal T}, {\cal Q}$, $\Theta$ (excepting ${\cal D}$ and ${\cal S}$) and $T'$ (pseudo-symmetric tridiagonal) all conform to the Poisson distribution $p_{\mu}(s) =\mu e^{-\mu s}$, where $\mu$ depends upon the choice of the matrix and PDF. Let  complex eigenvalues of $R$, $C$ and $T$ be $E^c_n$. We show that all $p(s)$ arising due to $\Re(E^c_n)$, $\Im(E^c_n)$ and $|E^c_n|$ of $R$, $C$ and $T$ are also of Poisson type: $\mu e^{-\mu s}$. So far the spacing distribution of mixed eigenvalues of an ensemble of real symmetric random matrix is known to be of Poisson type. We observe $p(s)$ as half-Gaussian for two real eigenvalues of $C$. Defining spacing as $S_k=|E^c_{k+1}-E^c_k|$ for real matrices $R, C, T$; when the complex eigenvalues are ordered by their real parts, we associate new types of $p(s)$ with them. Lastly, we study the distribution $D(\epsilon)$ of  eigenvalues of symmetric matrices (of large order) discussed above. For ${\cal R}$ and ${\cal R}'$, we recover semi-circle law and for $\cal C$ we get the known form. $\Theta$ gives a Gaussian and, ${\cal T}$ and $T'$ yield super-Gaussian distributions. For the secondary matrices ${\cal Q, D, S}$; $D(\epsilon)$ turns out to be exponential/sub-exponential.
	\end{abstract}	
\maketitle
\section{I. Introduction}
Due to matrix mechanics of Heisenberg and method of Linear Combination of Atomic Orbitals (LCAO) one can easily visualize the eigenspectrum of various systems with time-reversal symmetry as result of diagonalization of a real symmetric matrix where the matrix element are calculated using the inter-particle interaction. Most of the times this interaction is not known. For example   energy levels of various nuclei are known experimentally but the nuclear interaction is not really known.
Random Matrix Theory [1-5] originated by considering  the level spacing statistics $P(S)$ between eigenvalues of real symmetric matrices
\begin{equation}
R_1=\left (\begin{array}{cc} a & b\\ b & c\end{array}\right), \quad R_2= \left (\begin{array}
{cc} a+b & c \\ c & a-b \end{array}\right)
\end{equation}
when matrix elements $a,b,c$ are random numbers with the Probability Distribution Function (PDF) as 
Gaussian: $f(x)=\frac{1}{\sqrt{2\pi}} e^{-x^2}.$ The spacing of eigenvalues are given as ${\cal S}_1\sim\sqrt{4b^2+(a-c)^2}$ and ${\cal S}_2\sim \sqrt{b^2+c^2}$, respectively. Notice that the ${\cal S}_1$, is function of three parameters $(a,b,c)$, whereas ${\cal S}_2$ function of just two $(b,c)$. Notwithstanding this disparity and the complexity of the multiple integral 
\begin{small}
\begin{eqnarray}
P(S)=A\int_{-\infty}^{\infty}\int_{-\infty}^{\infty}\int_{-\infty}^{\infty} f(a,b,c) \delta[S-{\cal S}(a,b,c)]da~ db~dc, 
\end{eqnarray}
\end{small}
the spacing distributions in the two cases ($R_1,R_2$) turned out to be the same : $P(S)= S e^{-S^2}$ . When arranged to yield the average spacing as 1, the normalized spacing distributions is written as 
\begin{equation}
p_W(s)=\frac{\pi s}{2} e^{-\pi s^2/4}, 
\end{equation}
This is called the spacing distribution of Gaussian Orthogonal Ensemble (GOE) due to
the orthogonal symmetry of real symmetric matrices. Moreover $p(s)$ is well known as Wigner distribution function. Wigner surmised [1-5] that the spacing distribution of adjacent eigenvalues of  $N$ number of $n \times n$ Gaussian random real symmetric  matrices will again be given by (3). Next, Wigner predicted that Eq. (3) would eventually represent the the spacing statistics of neutron-nucleus scattering resonances and nuclear levels. Notice that near zero $p_W(s)$ is linear as $\pi s/2$, this is called the linear level repulsion of adjacent eigenvalues. The  rotational invariance and invariance under time-reversal of a symmetric matrix lie behind the linear level repulsion. Consequently, the spacing of nuclear levels of same angular momentum $J$ and parity $\pi$ indeed display [1-5] $p(s)$ in Eq. (3). Wigner's surmise is strange but true, each  nucleus behaves like a matrix of large order.
 
Wigner also conjectured  that $p(s)$ would not depend sensitively on the choice of $f(x)$.  By considering the probability distribution of matrix elements as uniform distribution and using a modest ensemble  of 200 real symmetric matrices of order $20 \times 20$, Rosenzweig's computations  have supported [6] Wigner's conjecture .

Even much later, in the recent years  investigations of spacing distributions of  ensembles random matrices  using $2 \times 2$ continue to be an attractive proposition for both  symmetric/Hermitian [7,8] and non-Hermitian matrices [9-14].
In these works in the Hermitian or real symmetric one has taken Gaussian distribution with zero mean and different variances for various entries of the matrices and derived a variety of spacing distributions [7,8]. Similarly, under Gaussian PDF,  for ensembles of several pseudo-symmetric, pseudo-Hermitian $2 \times 2$ representing Parity-Time-reversal symmetric systems novel expressions of $p(s)$ have been derived [9-14].

Here, we show that even two modifications of real symmetric $2 \times 2$ random matrices  yield different $p(s)$ under the same probability distribution function (PDF) $f(x)$. Similarly, one type of matrix under several non-Gaussian PDFs yield
 distinct expressions for $p(s)$. However, all the spacing distributions display the linear ($\alpha s$)  level repulsion near $s=0$ wherein $\alpha$ depends on the type PDF and the type of matrix (1) being used. The question of using a non-Gaussian probability distribution to test Wigner's second conjecture does not appear to have attracted attention after the Ref. [6]. However, use of many non-Gaussian distributions for finding the probability [15] of  occurrence of real eigenvalues for  the product of $n$ number of $2 \times 2$ real random matrices is worth mentioning.

Interestingly, though discussions of two modifications of real symmetric matrices (1) under several kinds of PDF of matrix-elements is like playing jeopardy with random matrices wherein the realization of (3) when both $n$ and $N$ are large appears to be far fetched, yet it happens giving us some surprises. 

The other interesting spectral statistics denoted as $D(\epsilon)$ is called distribution of eigenvalues of $n \times n$ real symmetric matrices when $n$ is large. Wigner proposed it to be [1-5]
\begin{equation}
D(\epsilon)= \frac{2}{\pi} \sqrt{1-\epsilon^2}, \quad \epsilon=E/E_*
\end{equation}
which is well known as Wigner's semicircle law. In case of large values of $n$, $E_*$ is the maximum eigenvalue of the matrix. Due to the historic connection of $2 \times 2$ matrices in RMT and specially due to the astonishing sameness of $p(s)$ in case of GOE for $n=2$ and $n>>2$,
the question arising here is as to what is the analytic form of $D(\epsilon)$ in case of $n=2$. This question has been overlooked in the past, however due to the numerical calculations of Porter
we know that qualitatively $D(\epsilon)$ makes an interesting transition from a bell shape to the semi-circle as the $n$ increases. In case of $n=2$, we collect a large number ($N$) of matrices and find the mean of positive eigenvalues to fix $E_*=E_m$ to obtain $D(\epsilon)$ both analytically and by finding their histograms numerically. The obtained $D(\epsilon)$ defy the Wigner's semi-circle law (4) (for $n=2$) and  our analytic/semi-analytic results agree excellently with the numerical histograms.

Further, in this paper for an ensemble of $N$, $n \times n$, we investigate whether real symmetry of matrices is sufficient to give rise to the spacing distribution of adjacent levels of a matrix as Wigner's surmise (3). We construct real matrix $R$ to construct a real symmetric ones as ${\cal R}=R+R^t$. The superscript $t$ denotes transpose of a matrix throughout in this paper.  Another version of real symmetric matrix can be created as ${\cal R}'_{ij}=r={\cal R}'_{ji}$, here $r$ is a random number. Next, we construct cyclic matrix $C$ [2] and tridiagonal matrix $T$ [15] and their symmetric counterparts as ${\cal C}$ [2] and ${\cal T}$ [16], respectively. 
Symmetric Toeplitz [2] $\Theta$ and non-symmetric (pseudo-symmetric) tri-diagonal matrix $T'$ having all eigenvalues as real have also been constructed. Next, we construct three secondary random symmetric matrices as ${\cal Q}=RR^t, {\cal D}=CC^t, {\cal S}=TT^t$. We study the distribution of spacing of adjacent levels of a matrix of the types discussed above for symmetric cases when eigenvalues $E_k$ are real. For non-symmetric matrices with complex eigenvalues $E^c_k$, we study the spacing between their real parts $\Re(E^c_k)$, imaginary parts $\Im (E^c_k)$ or their modulii $|E^c_k|$. By ordering the complex eigenvalues by their real parts we define the spacing of eigenvalues as $S_k=|E^c_{k+1}-E^c_k|$ to associate a level spacing distribution with non-symmetric matrices.

Earlier, Bose and Mitra [17] have derived 
\begin{equation}
D(\epsilon)=(\pi/4) |\epsilon| e^{-\pi \epsilon^2/4}, \quad \epsilon=E/\bar E,
\end{equation}
 for the ensemble of real cyclic symmetric matrices ${\cal C}$. We are able to produce this result excellently by numerical construction of histograms under several PDFs of matrix elements. An interesting spectral statistics which needs to called  First Adjacent Level Spacing (FALS) statistics is discussed in Refs. [18,19]. For ${\cal C}$ for $3\times 3$ case it has been obtained and shown to represent $p(s)$ for up to $15 \times 15$ cases [18]. However the $p(s)$ for $N$, $n\times n$ matrices when both $n$ and $N$ are large remains open. Similarly, for  $C$, FALS distribution between one real and the adjacent complex level has been found to display [19] linear level repulsion near $s=0$, irrespective of the order of matrices: $3 \times 3$ or $100 \times 100$. These two papers [18,19] also give several interesting connections  of cyclic matrices with physical problems. However, in the present our interest is to observe a robust $p(s)$ for complex eigenvalues of an ensemble of $N$, $n \times n$ cyclic matrices $C$ when both $n$ and $N$ are large. 

Excepting, the well known Wigner surmises (3,4) for real symmetric matrices ${\cal R}$ and ${\cal R}'$,
$p(s)$ for the ensemble of $N$, $n\times n$ ($n,N$ large) for  all other matrices discussed above is not known so far. For instance, for Gaussian ensemble of symmetric tridiagonal matrices the distribution of eigenvalues has been found to be different [20] from the semi-circle law but the level spacing distribution has not been discussed. We, in our pursuit of finding the robust behaviour of $p(s)$ for ensemble of various real random matrices mentioned above, we find that $N=1000$ and $n=100$ are  sufficient, results hardly change by further change in $N$ or $n$.

In Section II, we wish to present analytic or semi-analytic $p(s)$ for four non-Gaussian PDFs of elements of matrices for two types of $2 \times 2$. These non-Gaussian PDFs are : Uniform (U), Exponential (E: $f(x)= e^{-|x|}$, Super-Gaussian (SG: $f(x)=e^{-x^4}$) and Maxwellian (M: $f(x)=x e^{-x^2}$). In section III, we derive $D(\epsilon)$ for $R_1$ and $R_2$ and plot them in Fig. 4 the with their numerical histograms. In section IV, we discuss various methods of finding level spacing histograms of ensembles of 1000, 100 $\times$ 100 real symmetric and non-symmetric matrices discussed above. 
In section V, we discuss various matrices and resulting spectral statistics: $p(s)$ and $D(\epsilon)$. Some more PDFs used here  there are: Triangular [T: $f(|x|\le 1)=1-|x|, f(|x|>1)=0]$, Parabolic [P:$f(|x|\le 1)=1-x^2, f(|x|>1)=0]$  and Semi-circle $[S:f(|x|\le 1)=\sqrt{1-x^2}, f(|x|>1)=0]$. Asymmetric PDFs are half-Gaussian, half-uniform, etc. along with $[P_2: f(|x|<1)=(1-x^2)(1+x), f(|x|>1)=0]$ and $[P_3: f(|x|<1)=(1-x^2)(1+x)^2, f(|x|>1)=0]$. In section VI, we present our conclusions.

\begin{figure}[h]
	\centering
	\includegraphics[width=7 cm,height=5.cm]{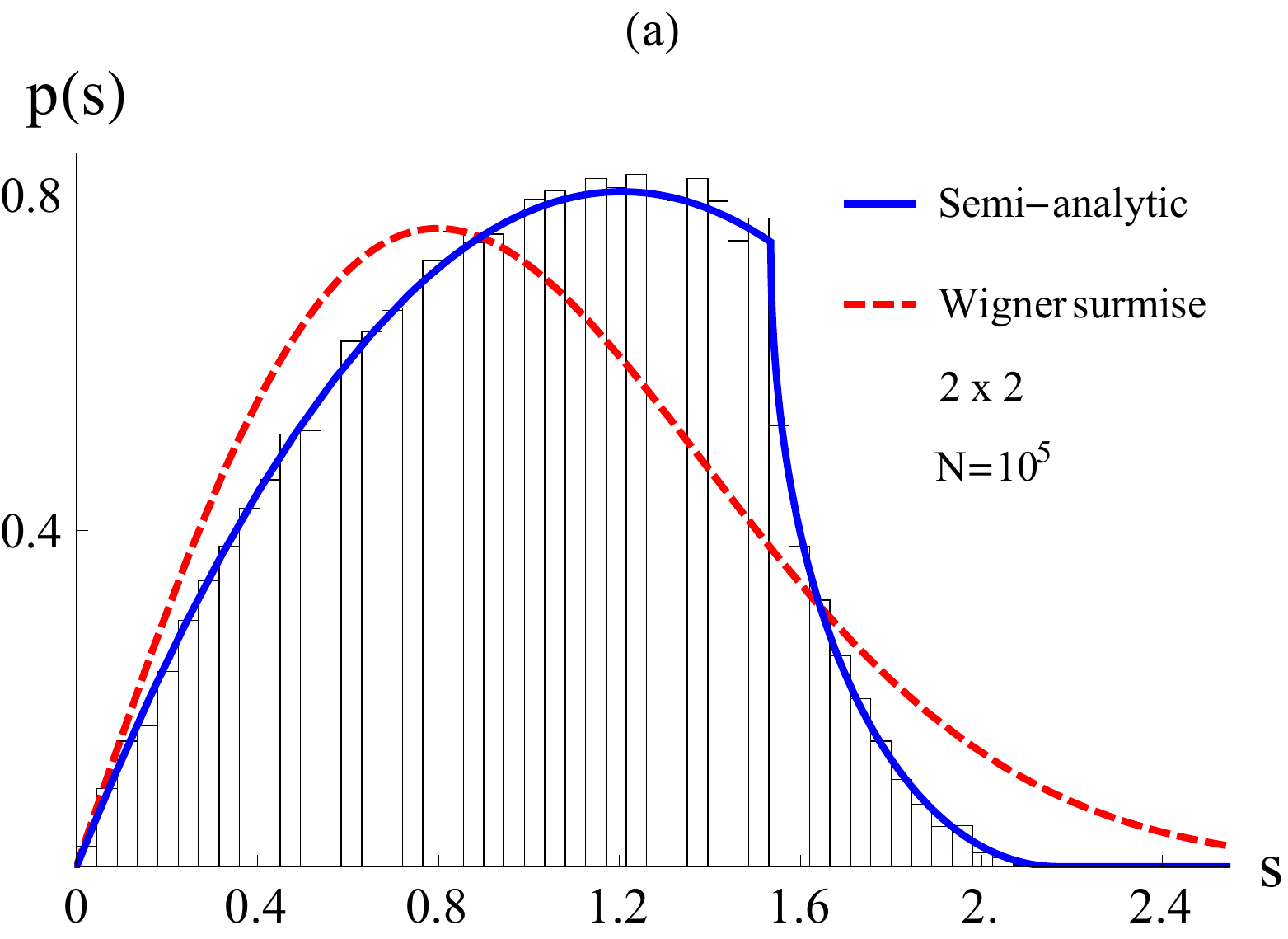}
	\includegraphics[width=7 cm,height=5.cm]{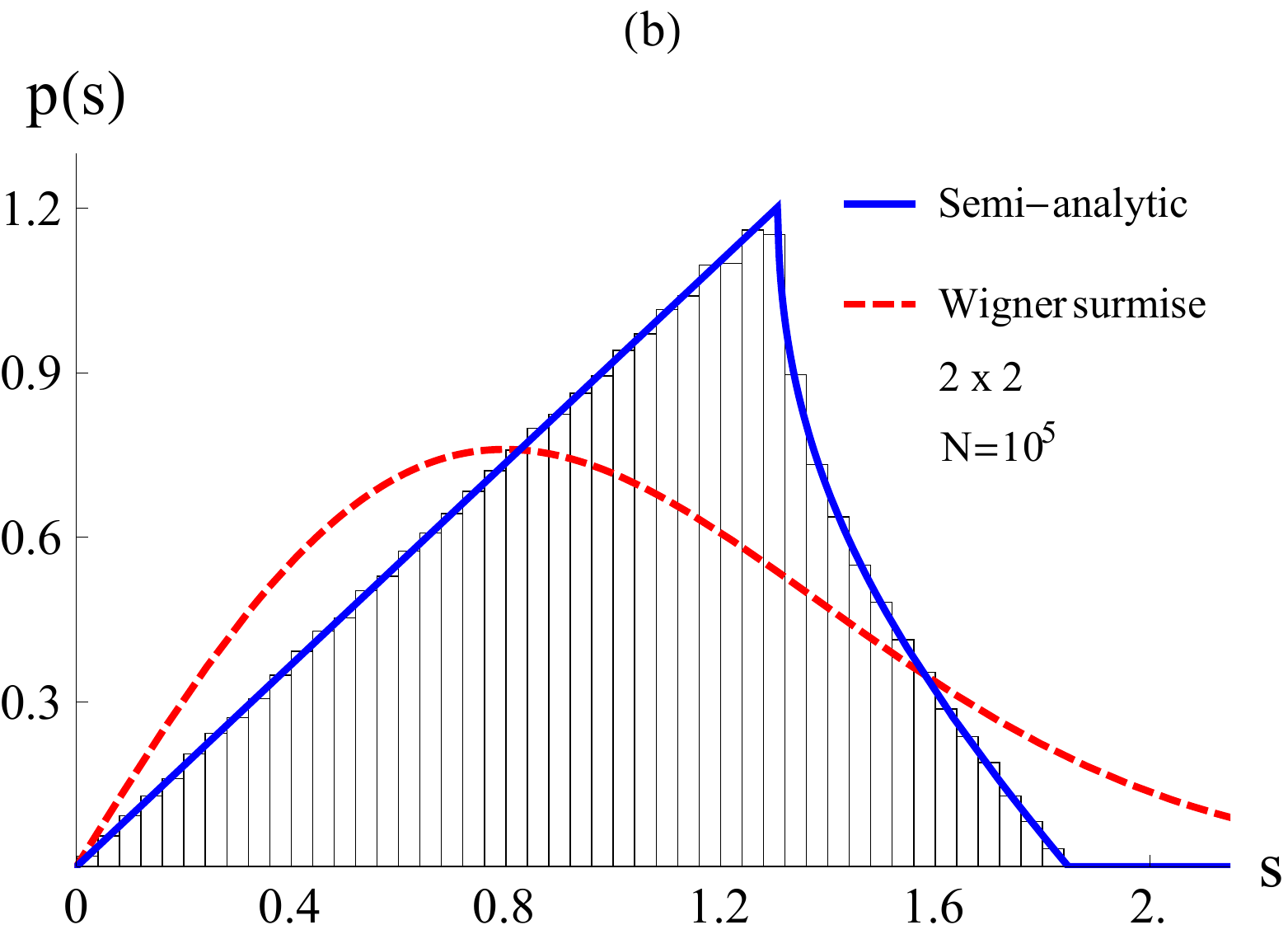}
	\caption{$p(s)$ for (a): $R_1$ and (b): $R_2$ matrices due to the uniform distribution of  elements. The solid lines are due to Eqs. (8, 11), dashed lines represent the  Wigner distribution (3). These two p(s) are distinctly different but near $s=0$ they are linear ($\alpha s)$, with $\alpha$ as 1.23 and 1.01, respectively}
\end{figure} 

\section{II. Ensembles of $2 \times 2$ real-symmetric random matrices and their spacing distributions}
In this section we find $P(S)$ (2) for two real symmetric matrices $R_1$ and $R_2$ for four  PDFs (U, E, SG, M). By finding the average spacing as $\bar S=\int_{0}^{\infty} S ~P(S)~dS/\int_{0}^{\infty} P(S)~dS$, we then find the normalized spacing distribution $p(s=S/\bar S)$. We thus conform to the invariance of distributions in two  $S$ and $s$ as $p(s) ds= P(S) dS$.

{\bf Uniform Distribution: $f(x)=1, 0<|x|\le \lambda, f(x)=0, |x|>\lambda$} 

For the matrix $R_1$, we have to evaluate
\begin{equation}
P(S)=A \int_{-\lambda}^{\lambda}\int_{-\lambda}^{\lambda}\int_{-\lambda}^{\lambda} \delta(S- \sqrt{4b^2+(a-c)^2})~ da~db~bc, 
\end{equation}
Without a loss of generality we may choose $\lambda=1$. Let us introduce the transformation from $(a,b,c)$ to $(u,v,w)$ as $u=a-c, v=a+c, w=2b$, Eq. (6) becomes
\begin{small}
\begin{eqnarray}
P(S)=\left \{ \begin{array}{lcr}
A \int_{0}^{S} du \int_{0}^{2} dw \int_{u-2} ^{2-u} \\ \delta(S-\sqrt{w^2+u^2})~dv & & 0 \le S <2 \\
A \int_{\sqrt{S^2-4}}^{2}~du \int_{0}^{2} dw \int_{u-2} ^{2-u} \\ \delta(S-\sqrt{w^2+u^2})~dv & & 2 \le S < 2\sqrt{2} \\ 
0 & & S.\ge 2 \sqrt{2}.\\
\end{array}
\right.
\end{eqnarray}
\end{small}

We find that  integrals in (7) can be done and $P(S)$ turns out to be a piecewise continuous function
given as
\begin{small}
\begin{eqnarray} 
P(S)=\left\{ \begin{array}{lcr}
A'~S(\pi-S)/4 & & 0\le S\le 2\\
A'~\frac{S}{2}[\sin^{-1}(2/S)
\\ -\sin^{-1}(\sqrt{S^2-4}/S)] \\ +\frac{S}{4}[\sqrt{S^2-4}-2],& & 2 <S<2\sqrt{2},\\
0, & & S \ge 2 \sqrt{2}. \\
\end{array}
\right.
\end{eqnarray}
\end{small}
For the matrix $R_2$ (1)
\begin{equation}
P(S)=A \int_{-\lambda}^{\lambda}\int_{-\lambda}^{\lambda}~\int_{-\lambda}^{\lambda} \delta[S-\sqrt{b^2+c^2}]~ da~db~dc.                                                                                                                                                                                                                                                                                                                                                                                                                                          	      
\end{equation}
The $a$-integral is separable and it will yield a multiplicative constant.
we convert the double integral in $b$ and $c$ in to polar form as $b=r \cos \theta, c=r \sin \theta$
\begin{small}
\begin{eqnarray}
P(S)=\left\{ \begin{array}{lcr}
A'\int_{0}^{1} \int_{0}^{\pi/4} r dr~ \delta(S-r) d \theta & & 0 \le S < 1\\
A'\int_{1}^{\sqrt{2}}\int_{\cos^{-1}(1/r)}^{\pi/4}
rdr\delta(S-r) d\theta & & 1\le S < \sqrt{2}\\
0. & & S \ge \sqrt{2}\\
\end{array}
\right.
\end{eqnarray}
\end{small}

Finally for real symmetric matrices (1) when the elements are distributed uniformly over [-1,1],
from (10) we get the continuous three piece spacing distribution function for $s \in (0, \infty)$ $(\lambda =1)$ as

\begin{equation} 
P(S)=\left\{ \begin{array}{lcr}
A'~\pi S/2, & & 0\le S\le 1\\
2A'~S~[\pi/4-\cos^{-1}(1/S)] , & & 1 <S\le\sqrt{2},\\
0, & & S > \sqrt{2}. \\
\end{array}
\right.
\end{equation}
In Fig. 1, we plot $p(s)$ arising from analytic results (8,11) along with the histograms generated from 100000 ($=N$), $2 \times 2$ real symmetric matrices of the types (a): $R_1$ and (b): $R_2$. Near $s=0$, they show linear repulsion, where $\alpha$ (the coefficient of linearity) is 1.23 and 1.09, respectively. Notice the excellent agreement of solid lines with histograms, the dashed lines represent Wigner's distribution (3).

\begin{figure}[t]
	\centering
	\includegraphics[width=7 cm,height=5.cm]{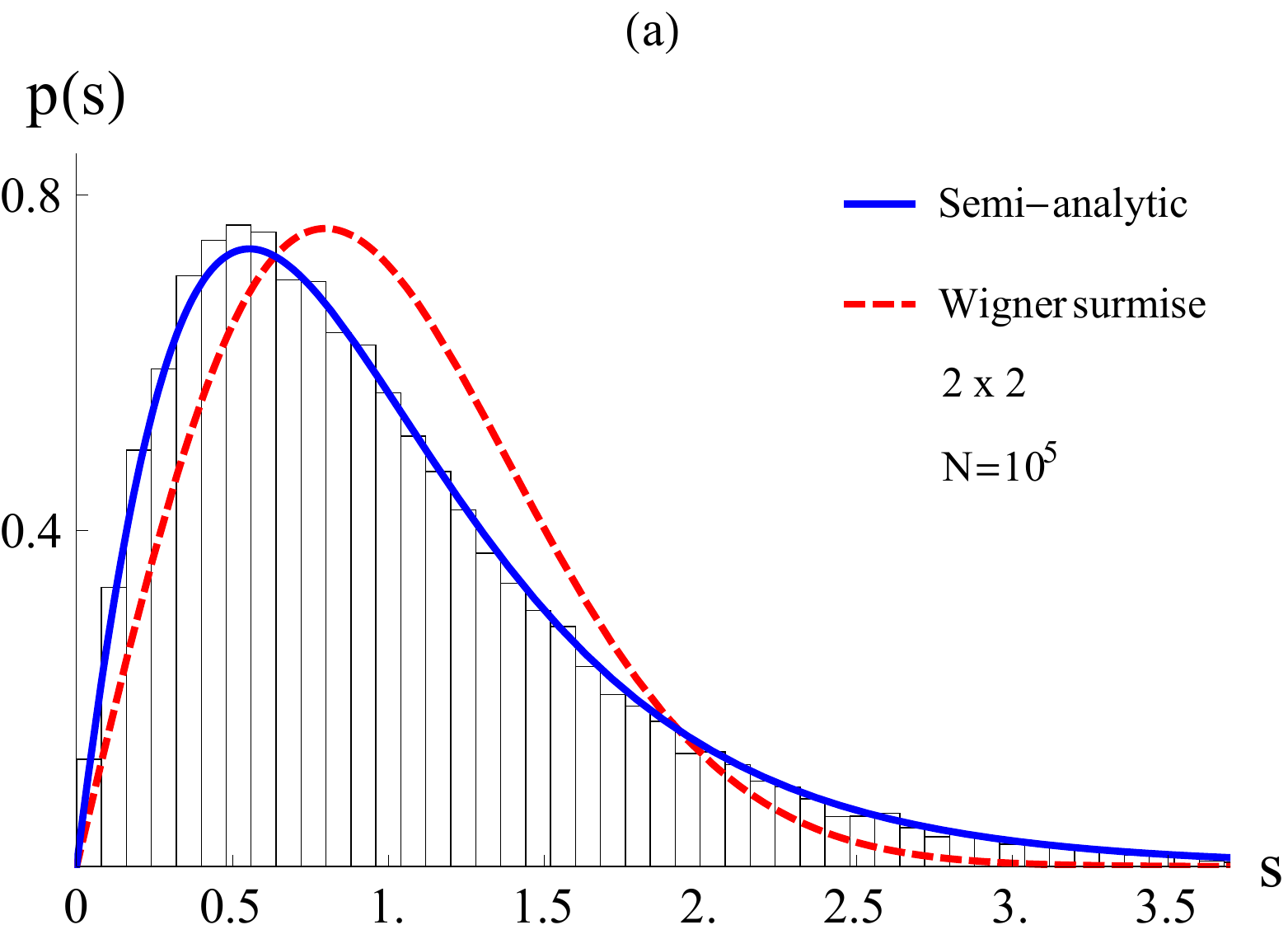}
	\includegraphics[width=7 cm,height=5.cm]{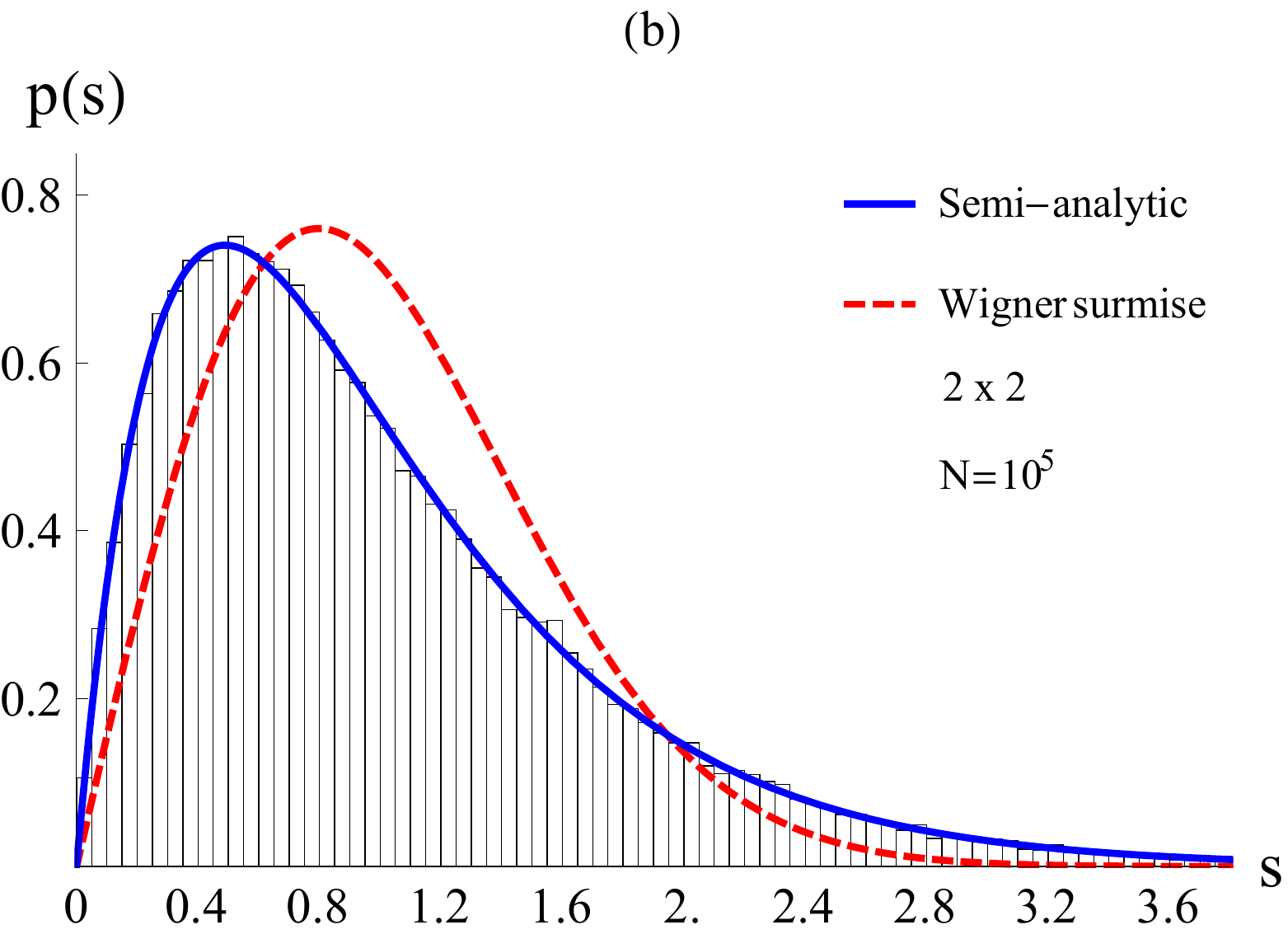}
	\caption{The same as in Fig. 1, for Exponential PDF (E: $e^{-|x|}$) arising from semi-analytic expression (14) and semi-analytic form (16). Here $\alpha$ values are 4.05 and 2.91, respectively.}
\end{figure} 

{\bf Exponential Distribution: $f(x)=e^{-|x|}$}

For exponential PDF, for $R_1$ we write 
the multiple integrals in $P(S)$ (2) can be made from 0 to $\infty$ and expressed as
\begin{small}
\begin{eqnarray}
P(S)=A \int_{-\infty}^{\infty}\int_{\infty}^{\infty}\int_{-\infty}^{\infty} e^{-(|a|+|b|+|c|)}  \\  \nonumber \delta[S-\sqrt{4b^2+(a-c)^2}]  da~db~bc. 
\end{eqnarray}
\end{small}
We transform $P(S)$ to the three dimensional the spherical polar co-ordinates using $2b=r \cos \theta,
a=r\sin \theta \cos \phi, c= r \sin \theta \sin \phi$ as
\begin{small}
\begin{eqnarray}
P(S)=A~ \int_{0}^{\infty} \int_{0}^{\pi} \int_{0}^{2\pi} e^{-r(cos\theta/2+\sin \theta (|\cos\phi|+|\sin\phi||)} \\ \nonumber \delta[S-r g(\theta, \phi)]~ r^2dr \sin \theta ~d\theta~ d\phi.
\end{eqnarray}
\end{small}
Crashing the delta function in above we get a $\theta, \phi$ integral
\begin{small}
\begin{eqnarray}
P(S)=A'\int_{0}^{\pi/2} \int_{0}^{\pi} e^{-S(\cos\theta/2+\sin \theta (|\cos \phi |+|\sin \phi|)/g(\theta, \phi)} \\ \nonumber \frac{S^2}{|g[\theta, \phi)|^3} \sin \theta ~d\theta~ d\phi,  ~g(\theta,\phi)=\sqrt{1-\sin^2\theta \sin 2 \phi},
\end{eqnarray}
\end{small}
Due to the symmetry of integrand the domains of integrations in (14) have been reduced. \\ \indent 
The $P(S)$ for $R_2$ for exponential distribution can be written as
\begin{small}
\begin{eqnarray}
P(S)=A \int_{-\infty}^{\infty}\int_{-\infty}^{\infty}\int_{-\infty}^{\infty} e^{-(|a|+|b|+|c|)}~\\ \nonumber \delta[S-\sqrt{b^2+c^2}]~ da~db~bc. 
\end{eqnarray}
\end{small}
Here the $a$-integral is separable and gives 1. The remaining double integral can be converted to polar form as
\begin{equation}	
P(S)=A'~S \int_{0}^{\pi/2} e^{-S(\sin \theta + \cos \theta)} ~d\theta. 
\end{equation}	
 These integrals are further  inexpressible in terms of known functions. $p(s)$ for these two cases are plotted in Fig. 2, they look similar though distinct, notice their linear behaviour near $s=0$  like Wigner's distribution (dashed line).
\begin{figure}[t]
	\centering
	\includegraphics[width=7 cm,height=5.cm]{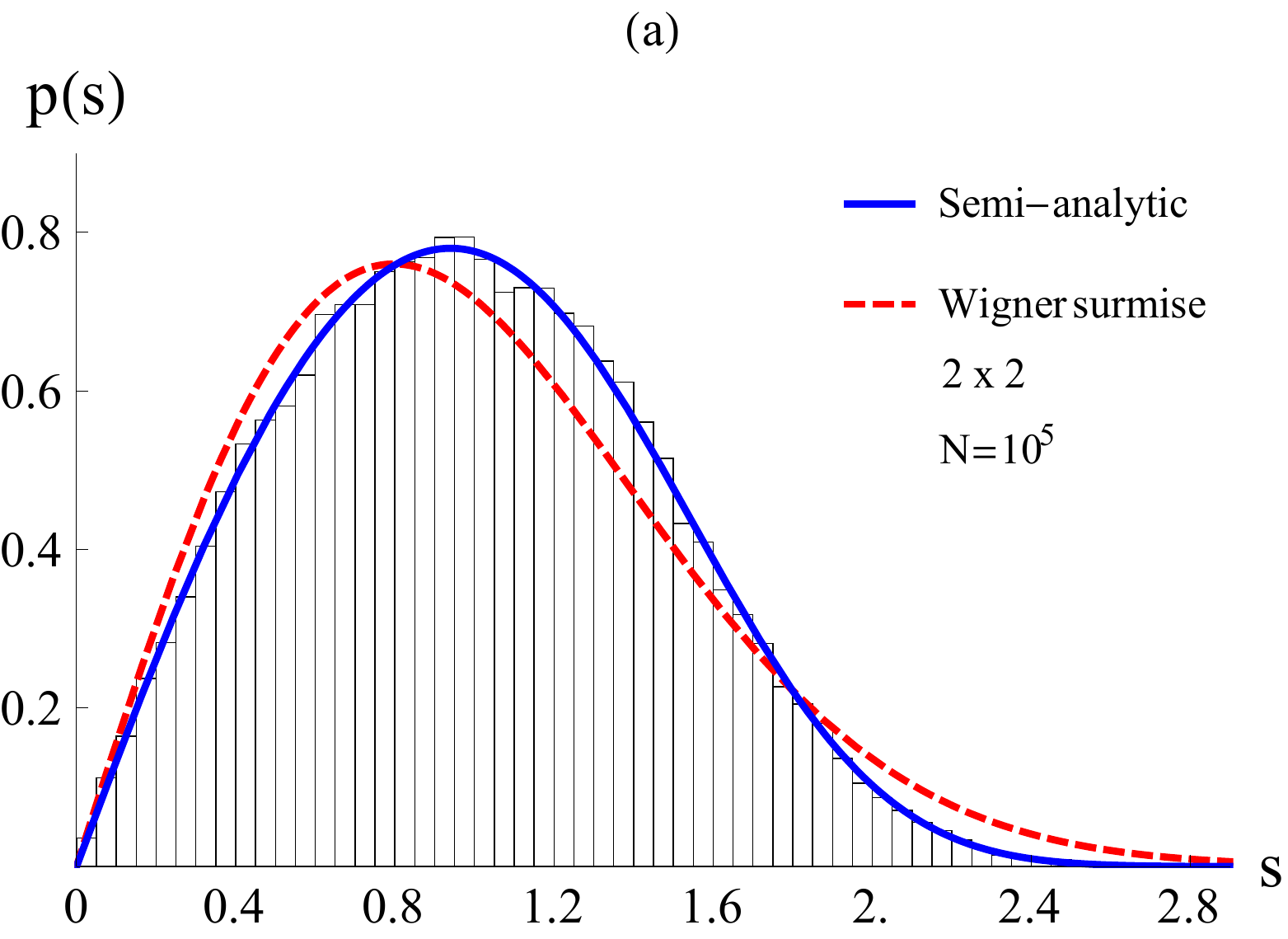}
	\includegraphics[width=7 cm,height=5.cm]{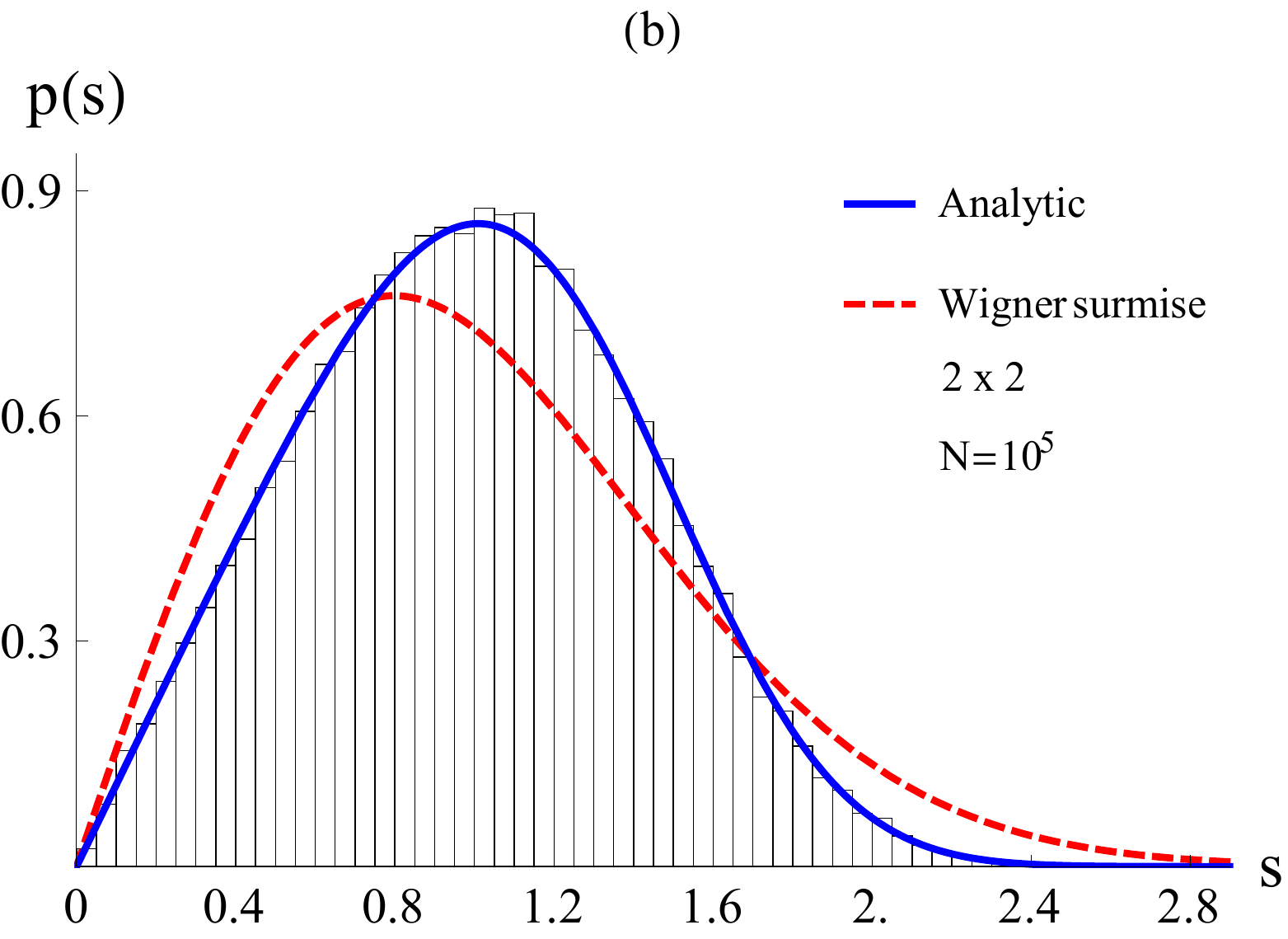}
	\caption{The same as in Figs. 1 and 2, for super-Gaussian PDF (SG: $e^{-x^4}$) arising from the
		semi-analytic expression (19) and the analytic one (20). Here $\alpha$ values are 1.30 and 0.91, respectively.}
\end{figure} 

{\bf Supper-Gaussian distribution: $f(x)=e^{-x^4}$}

For $R_1$ (1), the $P(S)$ integral (2) becomes
\begin{small}
\begin{eqnarray}
P(S)==A\int_{-\infty}^{\infty}\int_{-\infty}^{\infty}\int_{-\infty}^{\infty} e^{-(a^4+b^4+c^4)} \\ \nonumber \delta[S-\sqrt{4b^2+(a-c)^2}] ~da~db~bc. 
\end{eqnarray}
\end{small}
We transform $P(S)$ to the three dimensional the spherical polar co-ordinates using $2b=r \cos \theta,
a=r\sin \theta \cos \phi, c= r \sin \theta \sin \phi$ as
\begin{small}
\begin{eqnarray}
P(S)=A\int_{0}^{\infty} \int_{0}^{\pi} \int_{0}^{2\pi} e^{-r^4(cos^4\theta/16+\sin^4 \theta (\cos^4\phi+\sin^4\phi))}\\ \nonumber \delta[S-r g(\theta, \phi)]~ r^2dr \sin \theta ~d\theta~ d\phi.
\end{eqnarray}
\end{small}
Crashing the delta function in above we get a $\theta, \phi$ integral
\begin{small}
\begin{eqnarray}
P(S)=A'\int_{0}^{\pi/2} \int_{0}^{\pi} e^{-S^4(\cos^4\theta/16+\sin^4 \theta (\cos^4 \phi+\sin^4 \phi))/g^4(\theta, \phi)} \\ \nonumber \frac{S^2}{|g[\theta, \phi)|^3} \sin \theta ~d\theta~ d\phi,  g(\theta,\phi)=\sqrt{1-\sin^2\theta \sin 2 \phi},
\end{eqnarray}
\end{small}
where using the symmetry of integrand we can reduce the domains of integration as $(0,\pi/2)$ and $[0,\pi]$, respectively. \\ \indent 
For the matrix $R_2$, the $a$-integral in $P(S)$ is separable gives a multiplying constant. Then the double integral
in $b,c$ is changed to polar form where by crashing the delta function we get an integral
\begin{small}
\begin{eqnarray}	
P(S)=A'~S \int_{0}^{\pi/2 } e^{-S(\cos^4 \theta + \sin^4 \theta)} ~d\theta = A'~ S e^{-3S^4/4} \\ \nonumber \int_0^{2\pi} e^{-(S^4\cos t)/4}~dt  =\frac{A' \pi S}{2} e^{-3S^4/4}~ I_0(S^4/4),
\end{eqnarray}
\end{small}
$p(s)$ corresponding to (19) and (20) are plotted in Fig. 3 showing  linear level repulsion near $s=0$.

{\bf Maxwellian Distribution: $f(x)=x e^{-x^2}$}

For $R_1$ type of real symmetric matrix $P(S)$ is not simple, however here we would like to show that
when the PDF does not peak at $x=0$, we get highly non-linear behaviour of $P(S)$ near $S=0$ For $R_2$, To this end we convert the integral (2)  to polar form and get
\begin{equation}
P(S)=A' S^3 e^{-S^2},
\end{equation}
displaying nonlinear cubic behaviour $~S^3$ behaviour near $S=0$. 
\section{III Distribution of eigenvalues $D(\epsilon) $ of $2 \times 2$ Gaussian random matrices}
We collect $2N$ eigenvalues of $N$ $2\times 2$ matrices to find the mean of positive eigenvalue ($\bar E$) and divide all eigenvalues by $\bar E$ and find histograms $D(\epsilon)$. For a large real symmetric matrix this distribution is well known as semi-circle law.  

The distribution of eigenvalues  $E_1(a,b,c)$ and $E_2(a,b,c)$ can be obtained analytically  as 
\begin{small}
\begin{eqnarray}
g(E)= A\int_{-\infty}^{\infty} \int_{-\infty}^{\infty} \int_{-\infty}^{\infty}f(a,b,c) [\delta(E-E_1]+\delta(E-E_2)]\nonumber \\ da~db~dc, \bar E= \frac{\int_{0}^{\infty} g(E) dE}{\int_0^{\infty} g(E) dE}, \epsilon=\frac{E}{\bar E}, D(\epsilon)= \frac{g(\epsilon \bar E) }{\int_{-\infty}^{\infty} g(\epsilon \bar E) d\epsilon}. 
\end{eqnarray}
\end{small}
Once again $f(x)$ is the PDF of matrix elements.
Here, $E_{1,2}= \frac{1}{2}(a+c\pm\sqrt{(a-c)^2+4b^2})$ for $R_1$ and $E_{1,2}=a\pm\sqrt{b^2+c^2}$ for $R_2$.
For $R_1$, $g(E)$ can be obtained from Eq. (22), by using Gaussian PDF, defining $a+c=u, a-c=v$ and crashing the delta function w.r.t. $u$. Next, we use polar co-ordinates $v=r \cos\theta$ and $b=\frac{r}{2} \sin \theta$ to get 
\begin{equation}
g(E)=A' e^{-2E^2} \int_{0}^{\infty} \int_{0}^{2\pi}\cosh(2Er) e^{-[r^2(\frac{7}{8}+\frac{\cos 2\theta}{8}]} rdr d\theta
\end{equation}
which reduces to a one-dimensional integral
\begin{equation}
g(E)=A'' e^{-2E^2} \int_{0}^{\infty} r \cosh(2Er)~ e^{-7r^2/8} I_0(r^2/8)~ dr.
\end{equation}
For $R_2$ with Gaussian PDF, we crash the delta function w.r.t. the variable $a$ and use polar co-ordinates $b=r \cos \theta, c=r \sin \theta$ and we get a simple form
\begin{equation}
g(E)= e^{-E^2} [2 + \sqrt{2\pi} E ~ \mbox{erf}(E/\sqrt{2})~e^{E^2/2}]/(4\sqrt{\pi}).
\end{equation}
This function is normalized to 1 in for $E \in (-\infty,\infty)$, $\bar E$ calculated in $E \in (-\infty, \infty)$ is $\frac{4+\pi}{4\sqrt{\pi}} =1.0073,$ consequently, $D(\epsilon)=g(\epsilon)$. See the $D(\epsilon)$ histograms  in Fig. 4 for eigenvalues of $N=8 \times 10^4$ matrices:(a) $R_1$ and (b) $R_2$ where the matrix elements are Gaussian random numbers with mean 0 and variance 1. In Fig. 4, $D(\epsilon)$ (24,25) represent the histograms excellently. Usually, $D(\epsilon)$ is plotted by taking $\epsilon=E/E_{max}$
and $D(\epsilon)$ is studied for $-1 \le \epsilon \le 1$. With regard to this the x-axis could be scaled down to the domain $[-1,1]$ to see that the ensembles of $2 \times 2$ real matrices defy the semi-circle law which is observed for real symmetric matrices of large order. We also find that $D(\epsilon)$ for both $R_1$ and $R_2$ are sensitive to the PDF of matrix elements.
\begin{figure}[H]
	\centering
	\includegraphics[width=4.25 cm,height=4 cm]{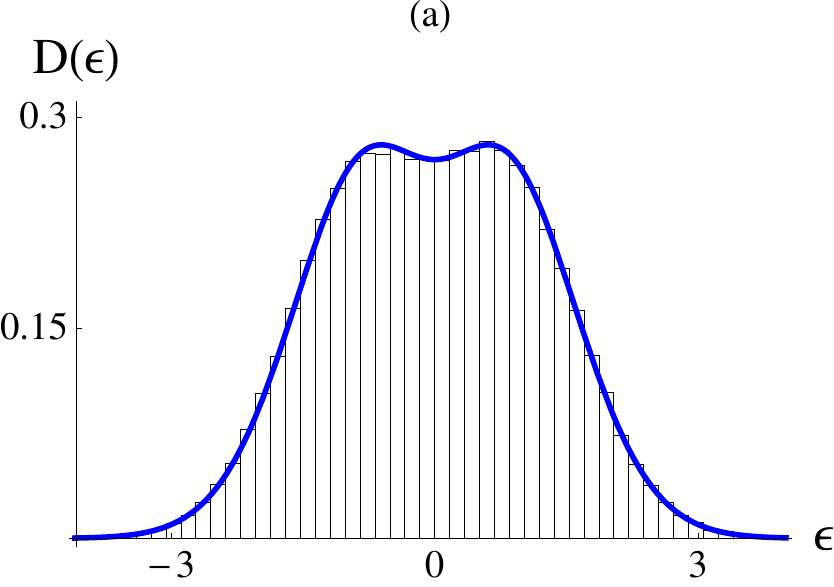}
	\includegraphics[width=4.25 cm,height=4.cm]{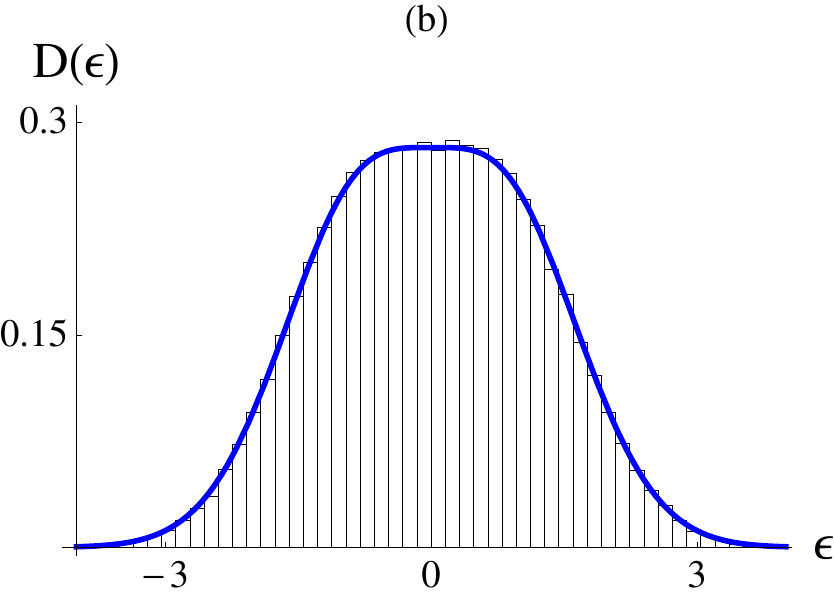}
	\caption{Distribution of eigenvalues $D(\epsilon)$ for $R_1$ (a) and $R_2$ (b) in (1) under Gaussian PDF of matrix elements. The solid blue line is due to Eqs. (24, 25). The histograms are due to an ensemble of $N=8\times10^4$ matrices.}
	\end{figure} 

\begin{figure}[h]
	\centering
	\includegraphics[width=7 cm,height=5.cm]{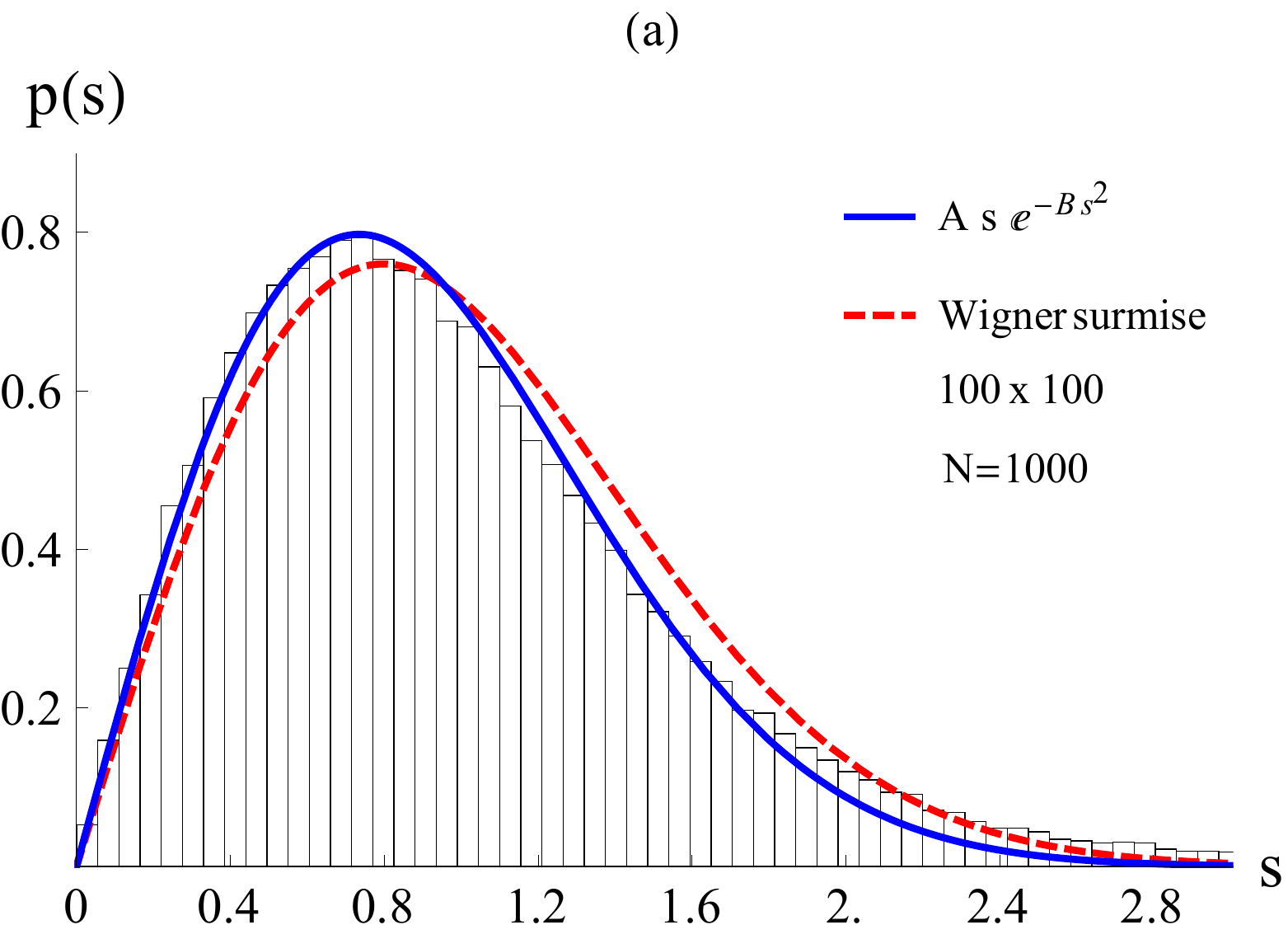}
	\includegraphics[width=7 cm,height=5.cm]{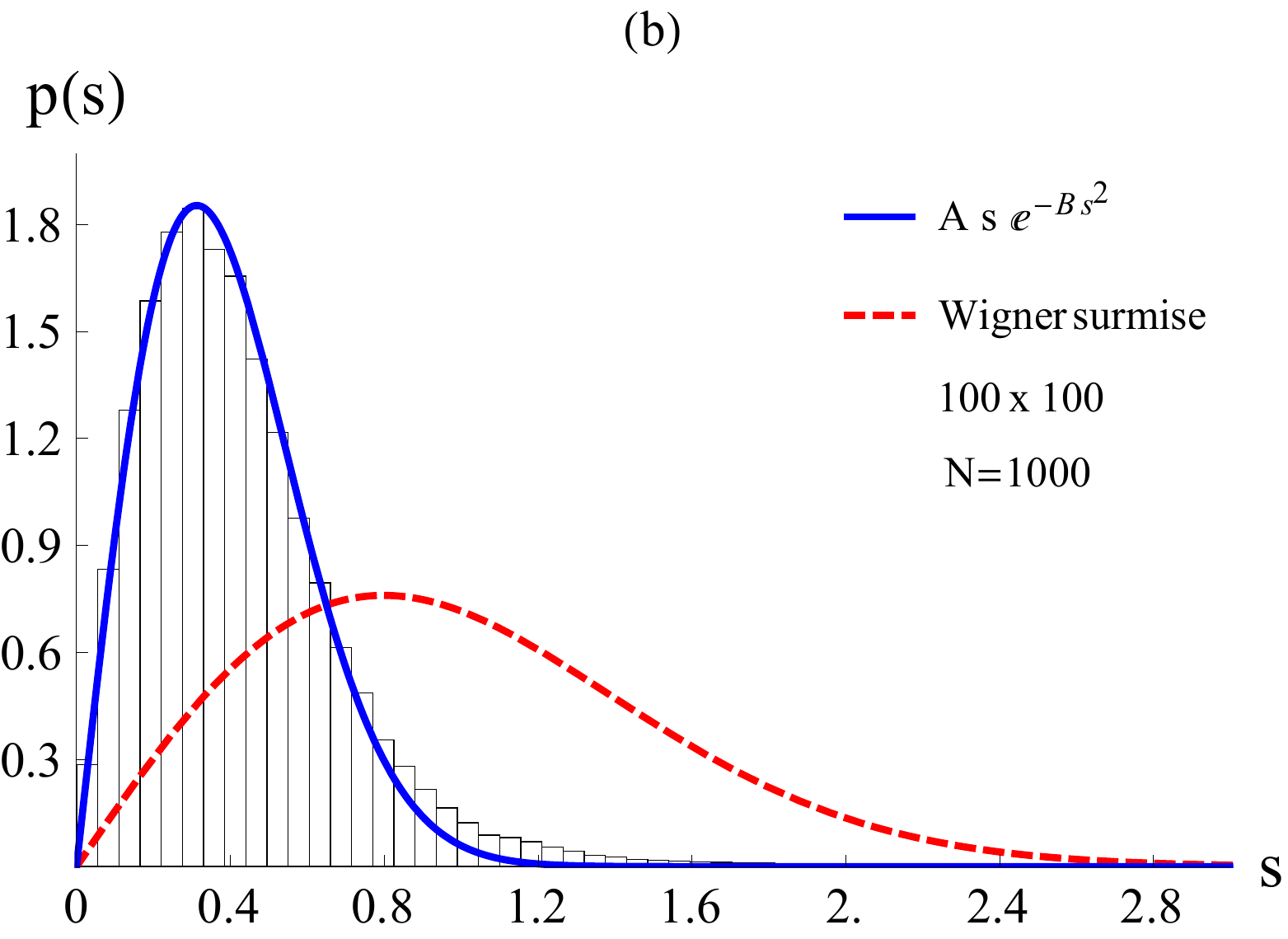}
	\includegraphics[width=7 cm,height=5.cm]{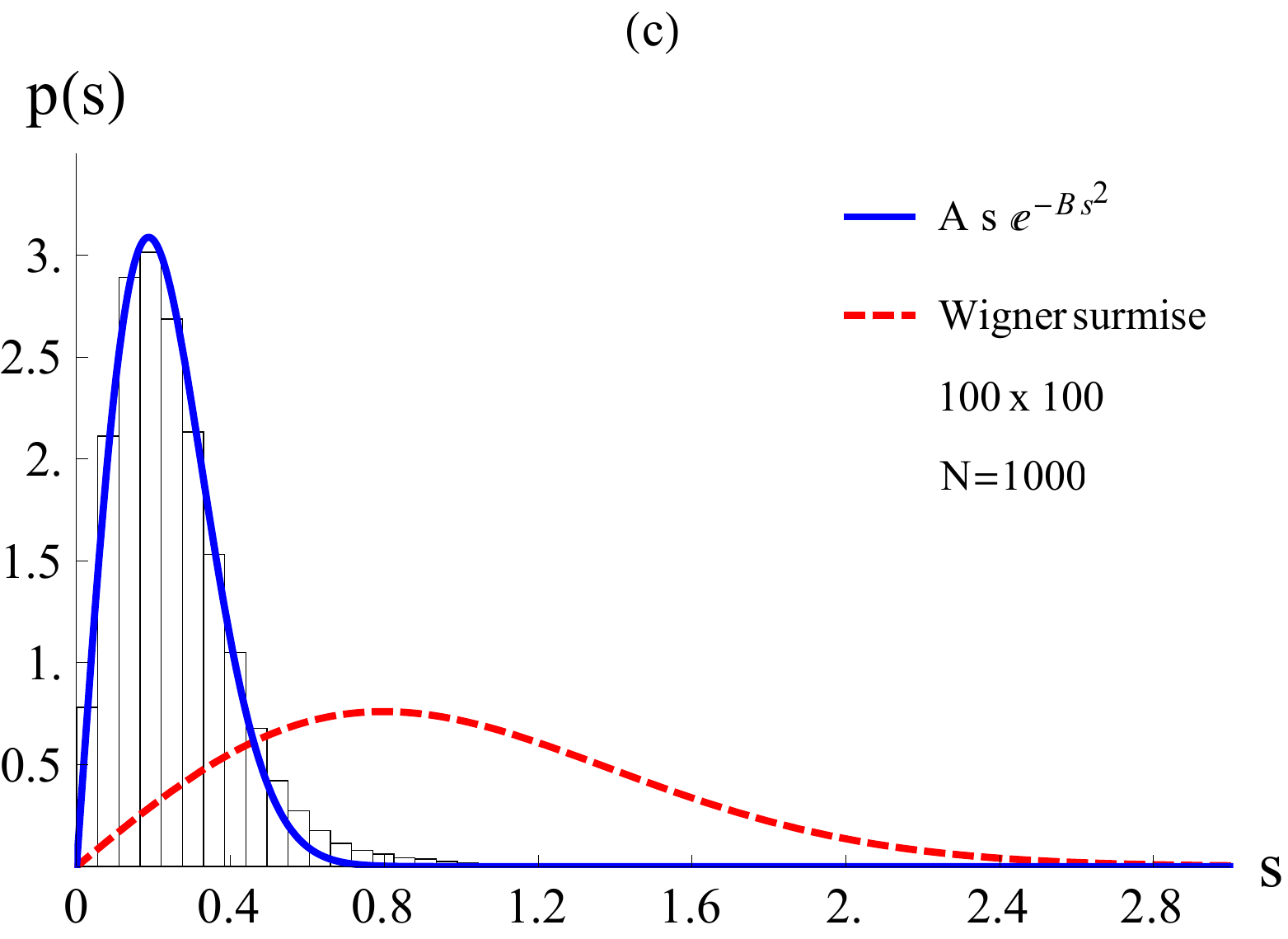}
	\caption{The transition of $p(s)$ in terms of fitted values of $A$ and $B$, when the PDF varies from symmetric to asymmetric. (a): for $f(x)=1-x^2:P$, $A=1.79, B=0.92$, (b): for $f(x)=(1-x^2)(1+x): P_2$, $A=9.72, B= 5.09$,(c): $f(x)=(1-x^2)(1+x)^2: P_3$, $A= 27.25, B=14.23$,  The domain of these $f(x)$ is $[-1,1]$ and $f(|x|>1)=0$. The solid lines show the fitted curve (26), dashed line is the Wigner distribution (3). Further increase in $N$ or $n$ values does not change this scenario. In part (a), notice that Wigner surmise succeeds only approximately for this non Gaussian PDF. These results are for ${\cal R}$ and ${\cal R}'$ see the Table. I, II.}
\end{figure} 

\begin{figure}[t]
	\centering
	\includegraphics[width=7 cm,height=5.cm]{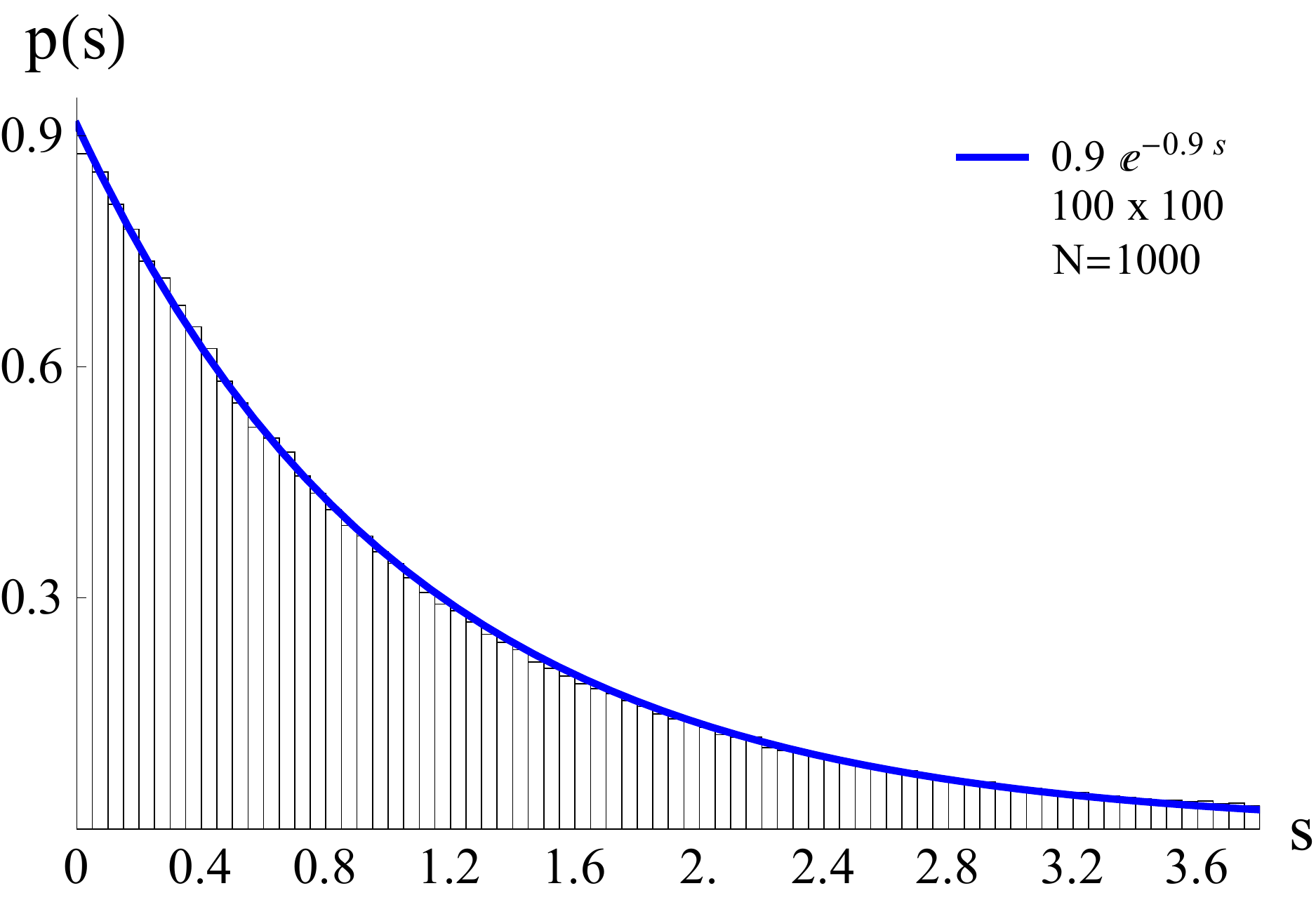}
	\caption{The good fit of histograms to $p_\mu(s)$ for $T$(tridiagonal), When the real part of $E_n^c$ has been used in NLM and the PDF of matrix elements is Gaussian. This is a typical scenario of good fit to $p_\mu(s)$ for the cases listed in Table-III and IV. Here $\mu = 0.90.$ }
\end{figure} 

\begin{figure}[h]
\centering
\includegraphics[width=4.27 cm,height=3.5cm]{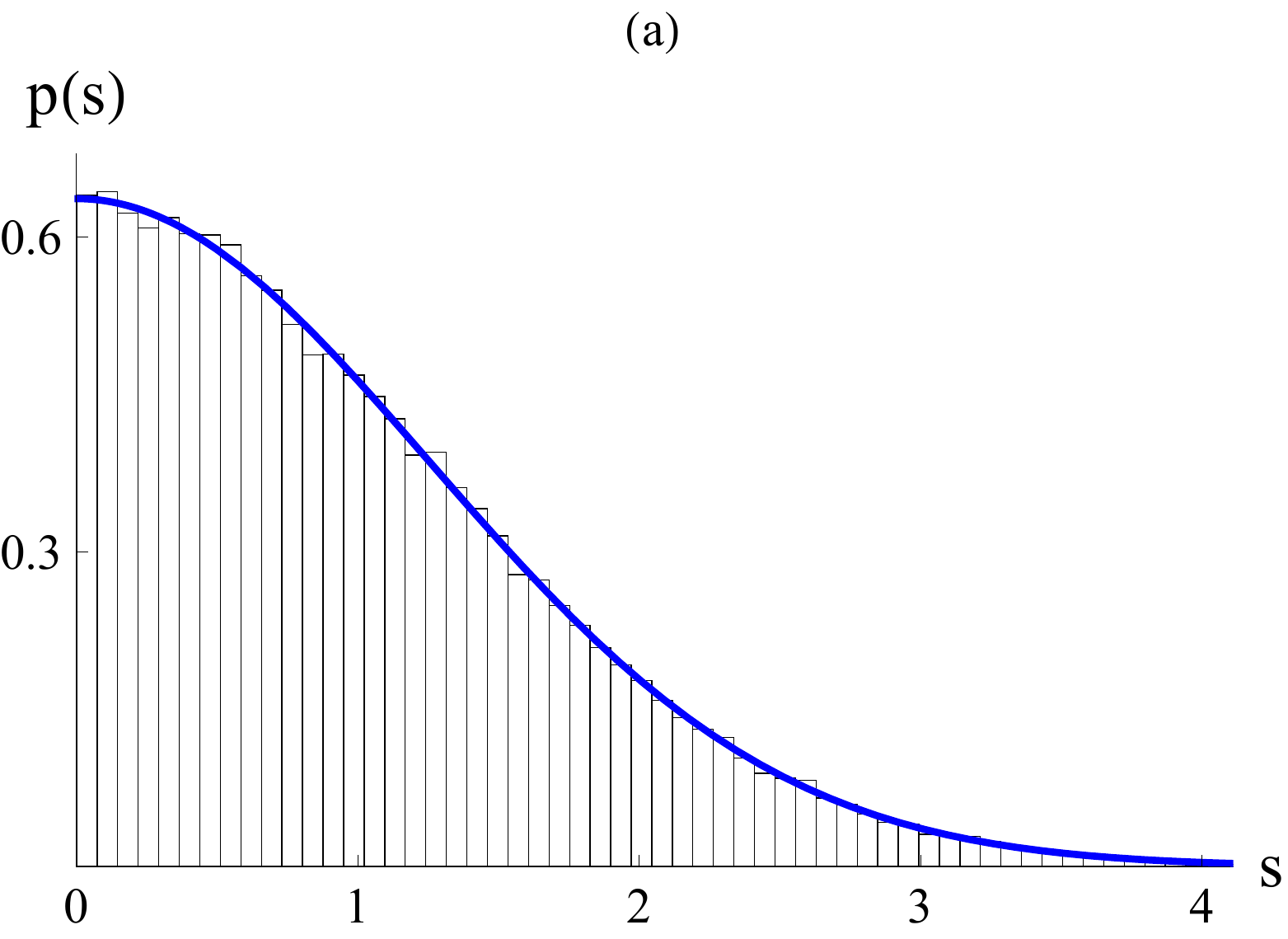}
\includegraphics[width=4.27 cm,height=3.5cm]{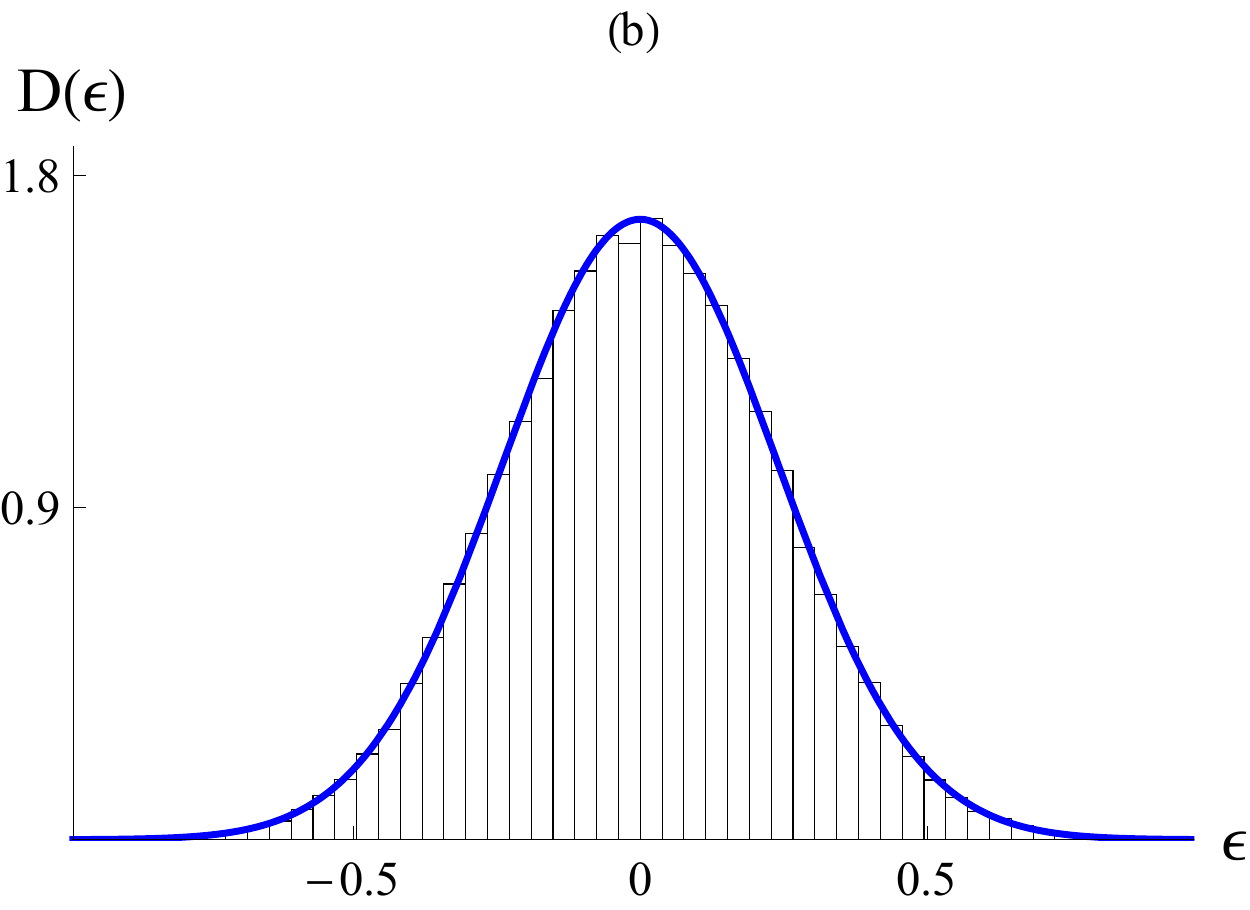}
\caption{$p(s)=\frac{2}{\pi} e^{-s^2/\pi}$ and $D(\epsilon)=1.68 e^{-8.72s^2}$ for two real eigenvalues of non-symmetric (pseudo-symmetric) cyclic matrices $C$, PDF of elements is Uniform distribution. We take $n=1000$ and $N=5000$ These result are also insensitive to the change of (symmetric) PDF.}
\end{figure} 
 
\section{III. Level spacing histograms of ensembles of N $n \times n$ real random matrices:}
\indent
In the literature [1-5] two types of methods have been discussed for find the nearest neighbour level spacing distributions of ensemble of $N$   $n \times n$ matrices. In the following, we out line them for our convenience and clarity naming them as NLE and NLM.\\

\indent{\bf Nearest Levels of Matrix (NLM) of ensemble:} We denote the statistics of the  spacings of adjacent levels of ensembles of $N$, $n \times n$  by $p(s)$ as in section II for $2 \times 2$ matrices. For these ensembles of $N$ matrices, for each $N$, we order $n$ eigenvalues in ascending order, find  spacings   $S_k=|E^c_{k+1}-E^c_k|, k=1,2,3,..n-1$. Let the average of $(n-1)$ values of $S_k$ be $\bar S$, we find  $N$ arrays of positive scaled $(n-1)$ spacings: $s_k=S_k/{\bar S}$. We then find histograms for these $(n-1)N$ numbers. One may also find the mean of all these  numbers and scale them again before finding histograms. Scaling brings the universality in $p(s)$ and one can compare them arising due to various PDFs of matrix elements. One may bypass one scaling of spacings or do both, $p(s)$ is known [1] to be insensitive to this choice. We re-confirm this for various PDFs. \\ 

\indent {\bf Nearest Levels of Ensemble (NLE) of matrices:} There is one  level statistics $p(s)$ which is done by collecting and ordering all the ($nN$) real eigenvalues to get $(nN-1)$ spacing of adjacent levels to do Histograms for the  spacings scaled (divided) by their global mean. $p(s)$ hence obtained is called level statistics of mixed eigenvalues. For a real symmetric matrix $p(s)$ is known to be following Poisson Distribution: $p(s)\sim e^{-s}$ , consequently the spacings of mixed eigenvalues are called un-correlated. 

\indent {\bf Level spacing  with complex energies:} In the case of non-symmetric real matrices some eigenvalues could be real but the rest of them are complex conjugate pairs, let us denote theM by $E^c_n$.  Since the number of the real eigenvalues may be arbitrary, in order to enhance the scope of Random Matrix Theory  we can order the complex eigenvalues by their real  parts to carry out the spacing statistics for $\Re(E^c_n) ,\Im(E^c_n)$ and $|E^c_n|$. One may define spacing as  $S_k=|E^c_{k+1}-E^c_k|$ and carry out spacing statistics for even complex eigenvalues by leaving the cases when $S_k=0$. In Ref. [19], authors propose  an interesting study of nearest level spacing
for eigenvalues which are `neither real nor complex conjugate pairs', more directly, we state
the nearest level spacing of complex eigenvalues $E^{c+}_k=a_k+i b_k$ taking only $b_k>0$ (or $b_k<0$). This would then correspond to study the complex level spacing in the upper (or lower) plane with $E^{c+}_k$ ordered by their real parts and $S_k$ defined as $|E^{c+}_{k+1}-E^{c+}_k|$.

\indent {\bf First Adjacent Level Spacing (FALS) distribution:} Recently, an interesting spectral statistics studied [18,19] could be called as First Adjacent Level Spacing distribution which determines the level spacings between first two real, one real and next complex or two nearest complex eigenvalues. Here, eigenvalues need to be ordered  by their real parts. 

\section{IV: Discussions of various real matrices and resulting $p(s)$ and $D(\epsilon)$}
The real symmetric matrices $n \times n$ matrices ${\cal R}$ and ${\cal R}'$ have $n(n-1)/2$ elements which are independent and identically distributed (iid) as per various PDFs. We find the good-fits to the histograms for $p(s)$ (NLM) for these two types of matrices by the function
\begin{equation}
p_{AB}(s)=A s e^{-Bs^2}
\end{equation}
for various (symmetric) PDFs. See Table I for the values of A and B, we observe that $A/2 \approx B \approx \pi/4$ hence they display Wigner's surmise approximately. The parameters A and B can be seen to be  almost insensitive to the choice of the matrix or the (symmetric) PDF. But in the cases of asymmetric PDFs (half-Gaussian etc.), we observe $A/2 \approx B >>\pi/4$. The values of A and B are sensitive to both the choice of the asymmetric PDF and the type of the matrix. In Fig. 5, we show the effect of asymmetry of PDF on $p(s)$. For symmetric case 5(a) Wigner's surmise is satisfied
(see entries for P in Table I), In Fig. 5(b,c) see the longish and squeezed distributions for  $P_1$
and $P_2$ (see Table II), where Wigner's surmise is not met.

Next, the $p(s)$ obtained by NLE by mixing the $1000(100\times 100)$ eigenvalues or ${\cal R}$ and ${\cal R}'$, we confirm [1] finding Poisson statistics $p_{\mu}(s)$. 
\begin{equation}
p_{\mu}(s)= \mu e^{-\mu s}.
\end{equation}
It is known that mixed eigenvalues of ensemble real symmetric matrices and hence the mixed nuclear levels do follow this statistics roughly as $e^{-s}$. The histograms found for the ensemble of 1000,
$100 \times 100$ for $Q$ matrices also turns out be (26).
The variation of $\mu$ w.r.t. various PDFs is given in Table III. Fig. 6, shows a typical fit of spacing histograms to $p_{\mu} (s)$. 

\begin{table}[]
	\centering
	\caption{Good fits of histograms of spacing distribution to $p_{AB}(s)$ (26) for 1000, 100 $\times$100  real symmetric random matrices (${\cal R}, {\cal R}'$)   for various PDF. Notice that $A/2 \approx B \approx \pi/2$.  The method NLM (Nearest Levels of Matrix) has been used. Compare  $A$ and $B$ values for PDF $P$(parabolic) to its asymmetric counterparts  $P_2$ and $P_3$ given in Table II. Also see Fig. 5(a,b,c).}
	\label{my-label} \vskip .5 cm
	\begin{tabular}{|c|c|c|c|c|}
		\hline
		\multirow{2}{*}{ $\hspace{0.05cm}$ $f(x)$  $\hspace{0.05cm}$ } & \multicolumn{2}{c|}{$\mathcal{R}$} & \multicolumn{2}{c|}{$\mathcal{R}'$} \\ \cline{2-5} 
		& $A$           & $B$         & $A$          & $B$          \\ \hline
		G                    & $\hspace{0.05cm}$  1.77  $\hspace{0.05cm}$      & $\hspace{0.05cm}$  0.93  $\hspace{0.05cm}$    &  $\hspace{0.05cm}$ 1.77  $\hspace{0.05cm}$      &  $\hspace{0.05cm}$ 0.93  $\hspace{0.05cm}$     \\ \hline
		U                    & 1.77       & 0.92     & 1.76      & 0.92      \\ \hline
		E                    & 1.75       & 0.93     & 1.77      & 0.93      \\ \hline
		SG                   & 1.76      & 0.92     & 1.77      & 0.92      \\ \hline
		T                    & 1.76       & 0.92     & 1.78      & 0.93      \\ \hline
		P                   & 1.79      & 0.93     & 1.77      & 0.92      \\ \hline
	\end{tabular}
\end{table}

\begin{table}[]
	\centering
	\caption{Values of the parameters $A$ and $B$ for fits of histograms to the spacing distribution $p_{AB}(s)$ (26) for ${\cal R}$ and ${\cal R}'$. The method NLM (nearest Levels of Matrix) has been used. Notice that $A/2 \approx B >> \pi/2$. These PDFs are $H_G:$ Half Gaussian, etc,  $P_2,P_3$ are other asymmetric PDFs, see in Fig. 5} \vskip .5 cm
	\label{my-label}
	\begin{tabular}{|c|c|c|c|c|}
		\hline
		\multirow{2}{*}{ $\hspace{0.05cm}$ $f(x)$  $\hspace{0.05cm}$ } & \multicolumn{2}{c|}{$\mathcal{R}$} & \multicolumn{2}{c|}{$\mathcal{R}'$} \\ \cline{2-5} 
		& $A$           & $B$         & $A$          & $B$          \\ \hline
		$H_G$                    & $\hspace{0.05cm}$  27.40  $\hspace{0.05cm}$      & $\hspace{0.05cm}$  14.41  $\hspace{0.05cm}$    &  $\hspace{0.05cm}$ 50.11  $\hspace{0.05cm}$      &  $\hspace{0.05cm}$ 26.23  $\hspace{0.05cm}$     \\ \hline
		$H_U$                    & 43.99       & 22.94     & 81.50      & 42.66      \\ \hline
		$H_E$                   & 30.13       & 15.89     & 17.15      & 09.09      \\ \hline
		$H_{SG}$                   & 36.23      & 18.94     & 67.25      & 35.16      \\ \hline
		$H_T$                    & 56.51       & 29.57     & 30.62      & 16.02      \\ \hline
		$P_2$                    & 09.72       & 05.09     & 05.96      & 03.12      \\ \hline
		$P_3$                    & 27.16       & 14.23     & 15.30      & 08.01      \\ \hline
	\end{tabular}
\end{table}

Next we construct two modifications of cyclic matrices: $C$ and ${\cal C}$ as:
\begin{eqnarray}
C=\left (\begin{array}{ccc} x_1 & x_2 & x_3\\ x_3 & x_1 & x_2 \\ x_2 & x_3 & x_1\end{array}\right), \quad {\mbox for}~ n \times n : C_{i,j}=x_k,\\ \nonumber k=(n+1-i+j)\mod 
n, \quad x_0=x_n,
\end{eqnarray}
which has complex conjugate eigenvalues excepting 1 and 2 real ones respectively for odd and even ordered matrices $C$. These matrices are also called circulants [2]. These eigenvalues are $\lambda_1=\sum_{k=1}^{n} x_k$ and $\lambda_{n/2+1}= \sum_{k=1}^n (-1)^k x_k$. We also point out that $C_{n \times n}$ is pseudo-symmetric as $\eta C \eta^{-1} =C^t$, where $\eta_{i,j}= \delta_{(i+j-1)\mod n,0}$. $\eta$ can be seen as generalized parity and $C$ being real, it can be called a PT-symmetric Hamiltonian [10-12]. Here $T$ stands for Time-reversals symmetry: $i \rightarrow -i$. Such Hamiltonians have either real or complex conjugate eigenvalues. The eigenvectors with real eigenvalues are orthogonal as $<\psi_1|\eta \psi_2>=0$ and eigenvectors corresponding to complex 
conjugate eigenvalues $E_{\pm}$ flip under parity as $\eta \psi_{+} =\psi_{-}$ and they are self-orthogonal as $<\psi_+| \eta \psi_+>= 0=<\psi_+|\psi_->$ [10-12]. For $2 \times 2$ Gaussian random pseudo-symmetric and pseudo-Hermitian matrices there are interesting results for the level spacing distribution. 
For the ensemble of $2\times 2$ pseudo-Hermitian matrices $p(s)$ has been show to be the same as Wigner surmise. However, for $n \times n$ ($n$-large) pseudo-Hermitian/symmetric matrices results are awaited
where the $p(s)$ due to  NLM arising from real eigenvalues are expected show a marked excursion from
Wigner's surmise. The histograms of $p(s)$ for two real eigenvalues of an ensemble of 1000, $100 \times 100$  cyclic matrices $C$ (27) with uniform distribution of its elements and its good fit to the normalized Gaussian  $p(s)=\frac{2}{\pi}e^{-s^2/\pi}$ in Fig. 7(a), may be seen as the first result in this regard. The distribution of there two real eigenvalues per matrix is shown in Fig. 7(b). We confirm the insensitivity of both $p(s)$ and $D(\epsilon)$ to the change of (symmetric) PDF of matrix elements. 

The level spacing distribution for two real eigenvalues of non-symmetric cyclic matrix $C$ is half-Gaussian (Fig.7a) irrespective of the type of (symmetric) PDF of matrix elements. This result fails the naive thinking that these two eigenvalues could be like $x_1+x_2$ and $x_1-x_2$ such that the spacing would be like $S= 2x_1, 2x_2$ so the distribution of $S$ may be expected as that of the PDF of the matrix elements $x_1,x_2.$

The second modification of the cyclic matrix is 
\begin{eqnarray}
{\cal C}=\left (\begin{array}{ccc} x_1 & x_2 & x_3\\ x_2 & x_3 & x_1 \\ x_3 & x_1 & x_2\end{array}\right), \quad {\mbox for}~ n \times n : {\cal C}_{i,j}=x_k,\\ \nonumber 
k=(i+j-1)~ \mod~ n, \quad x_0=x_n,
\end{eqnarray}
which is symmetric with all eigenvalues as real. $\hspace{-1cm} $This is also called persymmetric matrix [2] For 3$\times$3, case eigenvalues are $x_1+x_2+x_3, \pm \sqrt{[(x_1-x_2)^2+(x_2-x_3)^2+(x_3-x_1)^2]/2}$. Three ($s_0\hspace{-.15cm}=\hspace{-.1cm}\sqrt{2[(x_1-x_2)^2+(x_2-x_3)^2+(x_3-x_1)^2]}$, $s_{1,2}\hspace{-.18cm}=\hspace{-.0945cm}|x_1+x_2+x_3 \pm \sqrt{[(x_1-x_2)^2+(x_2-x_3)^2+(x_3-x_1)^2]/2}|$).
 In Ref. [18], the spacing statistics for partial spacings $s_0$ and
$s_{1,2}$ have been found which are also shown to fit well for the matrices up to order $15 \times 15$. In this paper, we are interested in the full spacing distribution (NLE) $p(s)$ for 1000 symmetric cyclic matrices ${\cal C}$ of order $100 \times 100$. Using various PDFs we find that
$p(s)$ for ${\cal C}$ is of Poisson type (26). The slight variation of $\mu$ is shown in Table III, where the PDFs have been changed. A typical fit
of  histograms to $p_{\mu} (s)$ is shown in Fig. 6. So far the spacing distribution of mixed ($nN$) nearest  levels of ensemble (NLE) of  real symmetric matrices is known to give rise to Poisson statistics (27). 

A matrix with the only nonzero entries on the main diagonal and the diagonals just above and below the main one is called tridiagonal matrix. We construct tridiagonal matrices as
\begin{small}
	\begin{eqnarray}
	T_{i,j}=\left\{ \begin{array}{lcr}
	x_i & & i=j \\
	y_i & & i=j-1 \\
	z_{i-1}. & & i=j+1 \\
    0. & &  |i-j| > 1 \\
	\end{array}
	\right.
	\end{eqnarray}
\end{small}
Where the real random numbers $x_k,  y_k,  z_k$ are $3n-2$ iids under a PDF. This real matrix $T$ has either real or complex conjugate eigenvalues. When $y_k = z_k$, the matrix becomes symmetric tridiagonal ${\cal T}$ having $2n-1$ iids and its eigenvalues will be real. When $ y_k z_k > 0$, it will be non-symmetric (pseudo-symmetric) $T'$ having all eigenvalues as real. Interesting features of tridiagonal matrices can be seen in Ref. [16]. Tridiagonal matrices are often encountered in quantum mechanics when one seeks the solution Schr{\"o}dinger equation by Frobinus method or for finding eigenvalues for some interesting potentials.

A matrix where the diagonal element is fixed is called Toeplitz matrix [2]. We construct real symmetric Toeplitz matrix $\Theta$ as, 
\begin{equation}
\Theta_ {i,j}= x_{|i-j|+1}
\end{equation}
Where the random numbers  $x_k, k \in(1,n)$ are $n$ iids under a PDF. In $\Theta$, all the diagonal elements are fixed as $x_1$ and all the eigenvalues are real. By considering ensembles of 1000, 100 $\times$ 100 matrices of the type ${\cal T}$, $ T'$ and $\Theta$ we construct histograms of nearest levels spacings using NLM. Poisson distribution (27) again fits  these histograms well like in Fig. 6. The variation of $\mu$ with respect to change of PDF is displayed in Table III.

 \indent In Table IV, we present the values of $\mu$ by good fits of  histograms to $p_{\mu}(s)$   for real matrices $R$, $C$ and $T$ which have complex eigenvalues as $E^c_n$, in these cases, $p(s)$ have been obtained by using $\Re(E^c_n)$, $\Im (E^c_n)$ and $|E^c_n|$ in NLM. In all of these cases where we take 1000, $100 \times 100$ matrices, we find that $p(s)$ is Poisson type: $p_{\mu}(s)$ (27). So $p_{\mu}$ can be seen insignificantly sensitive to various PDFs. However, for a fixed PDF they are significantly sensitive to the type of the matrix. Fig. 5 displays a typical fit of $p_{\mu}(s)$ to the histograms. 
 
For the real symmetric matrices ${\cal D}=CC^t$ and ${\cal S}= TT^t$, the level spacings histograms
for ensemble of 1000, $100 \times 100$ using NLM turn out to be a new type under any PDF. In fig. 7,
we present them for ${\cal D}$ under the Gaussian PDF. The real symmetric matrices ${\cal D}=CC^t$ and ${\cal S}=TT^t$ due to their less than $n^2/2$ correlated matrix elements display a new kind of sub-exponential statistics
\begin{eqnarray}
p_{ab}(s)=(a^{1/b}/\Gamma[1+1/b])~ e^{-as^b}, \quad 0<b<1.
\end{eqnarray}
For the case of ${\cal D}$ the parameters are $a=2,b=1/2$ and for ${\cal S}$ they are $a=4.6, b=0.3$
see Fig. 8. Here the elementary matrices $C$ and $T$ have $n$ and $3n-2$ (both $<< n^2$) number of iid matrix elements. Remarkably, for ${\cal Q}=RR^t$ due to $R$ matrix having $n^2$ iid matrix elements the spacing distribution is distinctly different (an exponential), see Table III..

Though Wigner's surmise through its $p_W(s)$ (3) has met success only for real matrices ${\cal R}$ and ${\cal R}'$, the  non-occurrence of the same p(s) for every other symmetric matrices discussed above makes the Wigner's surmise even more special. Now let us study $p(s)$ for non-symmetric matrices $C, T$ and $R$ by obtaining histograms for the ensemble of 1000, $100 \times 100$ using
$S_k=|E^c_{k+1}-E^c_k|$ in NLM. See Fig. 8, for $C$ with uniform distribution of matrix elements.
This is a non-standard $p(s)$ with good fit as $1.18 (s+.28) e^{-(s-0.39)^2/2}$. For the tridiagonal matrix $T$ under uniform distribution we find good fit to histograms as $p(s)=0.08(s+7.8)e^{-(s-0.68)^2/0.82}$.
Fig. 9, presents, the histograms for the matrix $R$ yielding a different kind of $p(s)$. These results are not sensitive to further increase in $N$, $n$ and change of PDF.
\begin{figure}[H]
	\centering
	\includegraphics[width=7 cm,height=5.cm]{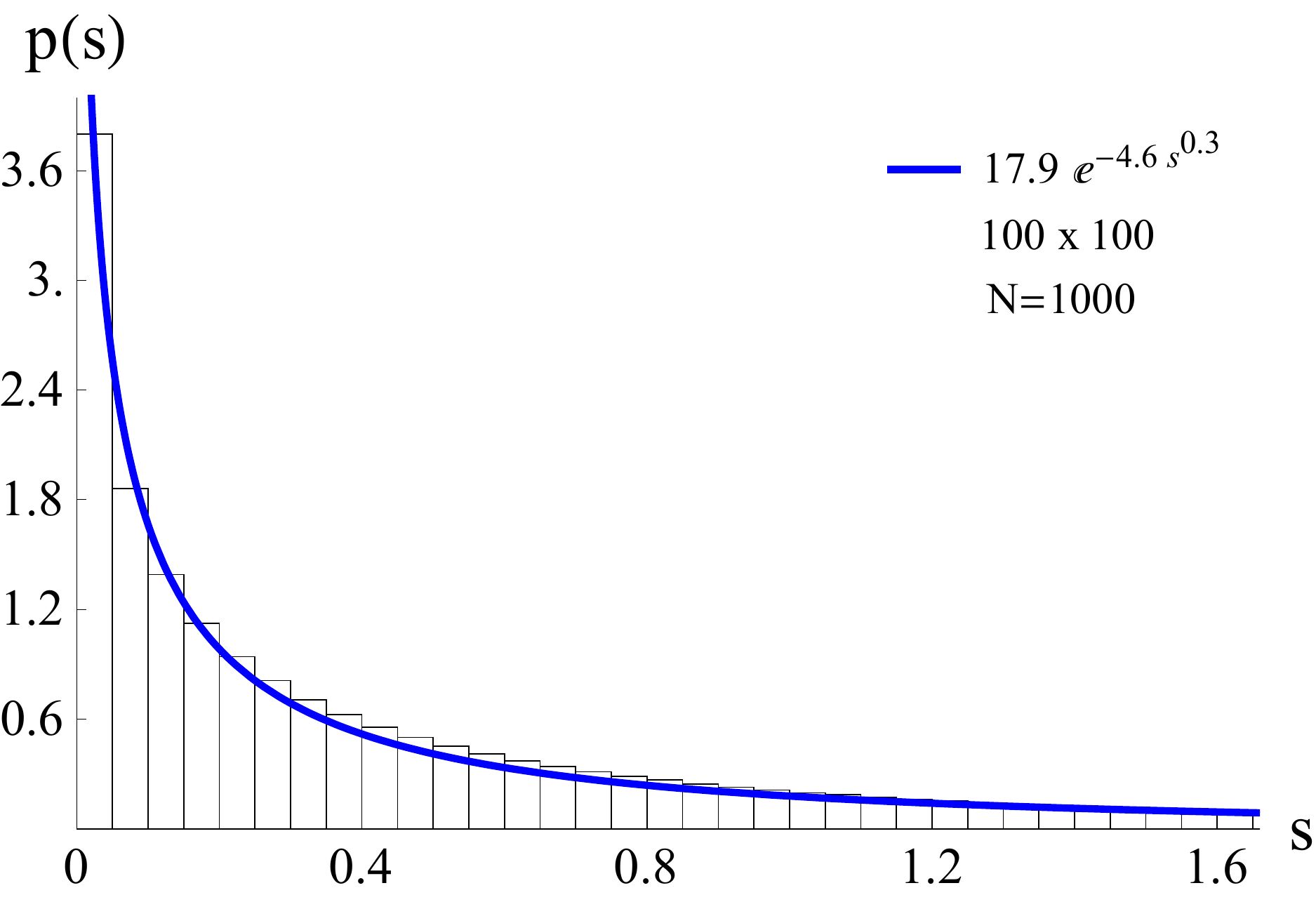}
\caption{sub-exponential $p(s)$ (see Eq. (32)) for the real eigenvalues of the symmetric matrix ${\cal S}=TT^t$. The PDF of elements of $T$ is uniform distribution. This result is  insensitive to the change of (symmetric) PDF. We get similar sub-exponential $p(s)$ for ${\cal D}=CC^t$, see the text below Eq. (32).}
\end{figure}

\begin{figure}[h]
	\centering
	\includegraphics[width=7 cm,height=5.cm]{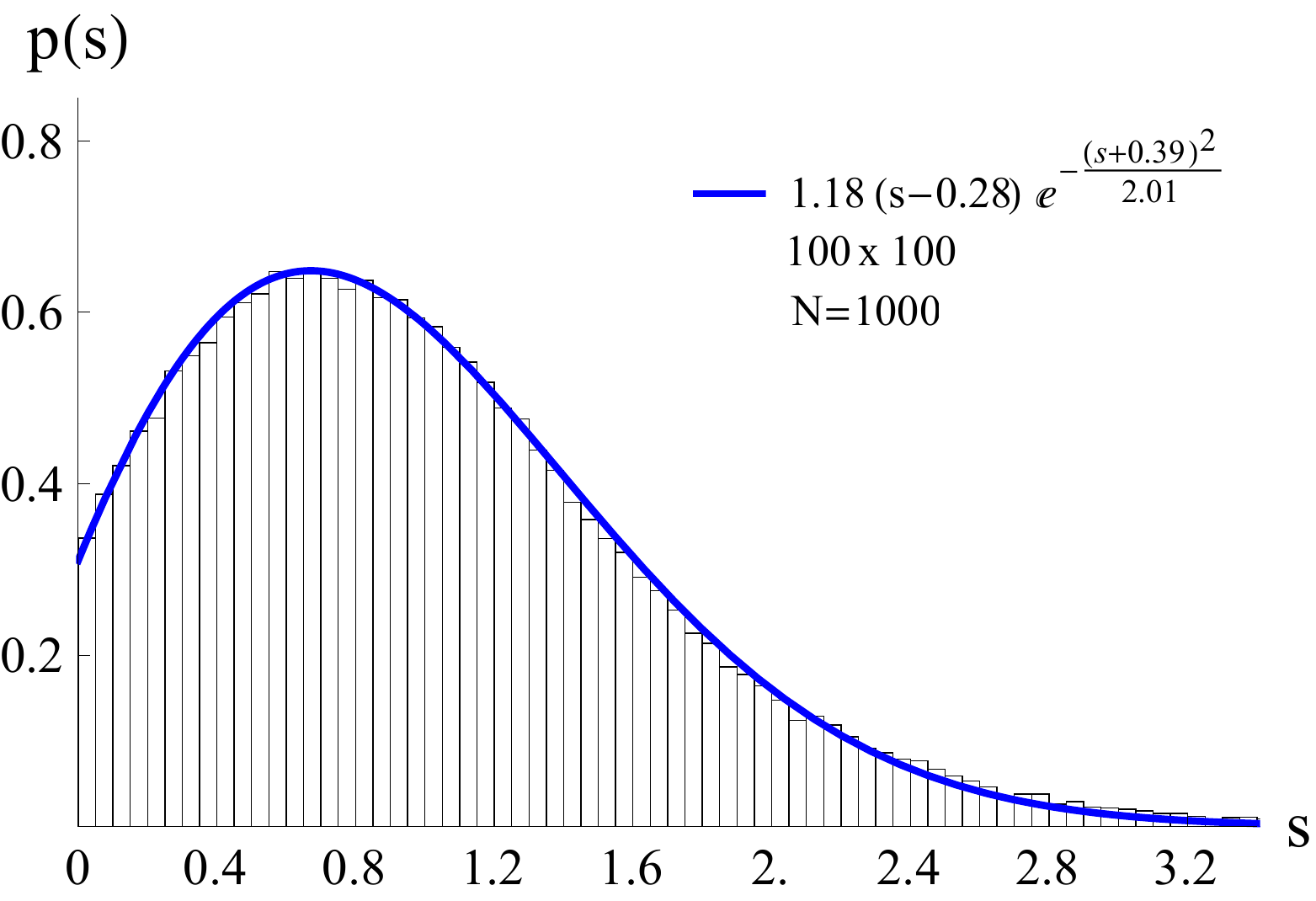}
	\includegraphics[width=7 cm,height=5.cm]{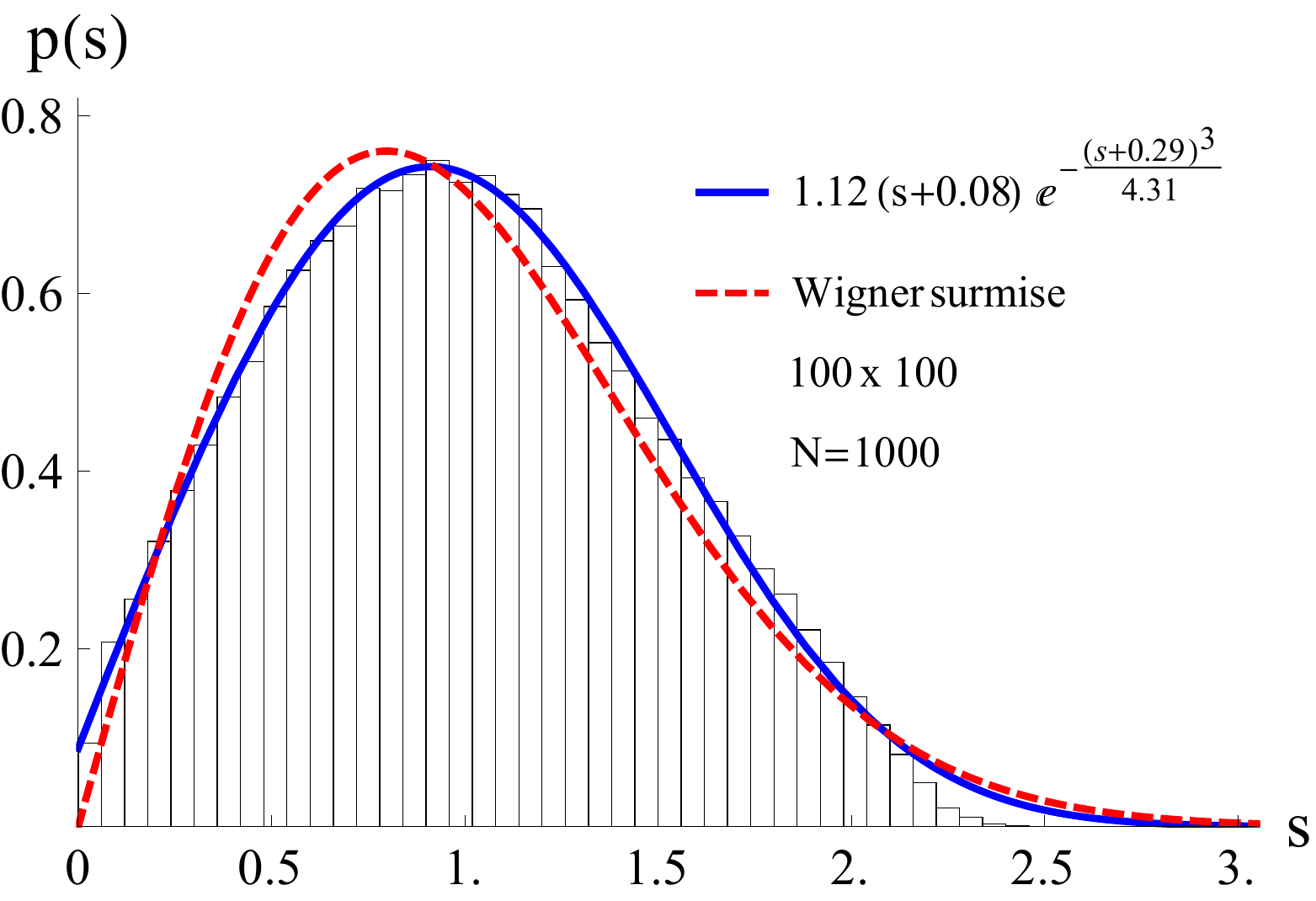}
	\caption{(a): $p(s)$ for  non-symmetric cyclic matrices  $C$, where PDF of elements is uniform distribution.  Here the eigenvalues are in general complex and they are ordered by their real part, and spacing between adjacent levels is defined as $ S_{k} =|E^c_{k+1}- E^c_{k}|$, We get a similar fit to the histograms for non-symmetric tridiagonal matrix $T$, where the fitted $p(s)=0.08(s+7.8)e^{-(s-0.68)^2/0.82}$. These results are insensitive to the change of (symmetric) PDF.(b): The same as in (a) for the non-symmetric real matrix $R$.}
\end{figure} 
 
Recently, there is an interesting attempt [19] to associate linear repulsion near $s=0$ or $p_W(s)$ with non-symmetric cyclic matrices having complex eigenvalues. In this regard, they find level spacing statistics  by calculating FALS distribution. We confirm their finding that the level repulsion near $s=0$ is exactly same as $\pi s/2$ irrespective of $C$ being $3 \times 3$ or $100 \times 100$ (see Fig. 2 of [19]). However, we do not find the spacing statistics of complex
eigenvalues of $C$ in the upper plane to be  the Wigner's $p_W(s)$ (3) as claimed (in Eqs. (14-15) and Fig. 3 of Ref. [19]). Instead, we get the empirical fit to $p(s)$ for $C$ and $T$ commonly as $p(s)=a s e^{-bs^2}, 0<b<1$ and for $R$ we get $p(s)= a s^b e^{-cs^d}, 0<d<1.$ (see Fig. 10(a))
Note that in the case of $R$ which has $n^2$ iid elements, $p(s)$ has a different form which near $s=0$ goes roughly as $s^3$ (see Fig. 10(b)). Also see the deviation of these new $p(s)$ from Wigner's $p_W(s)$ which is plotted by dashed lines.

\begin{figure}[h]
	\centering
	\includegraphics[width=7 cm,height=5.cm]{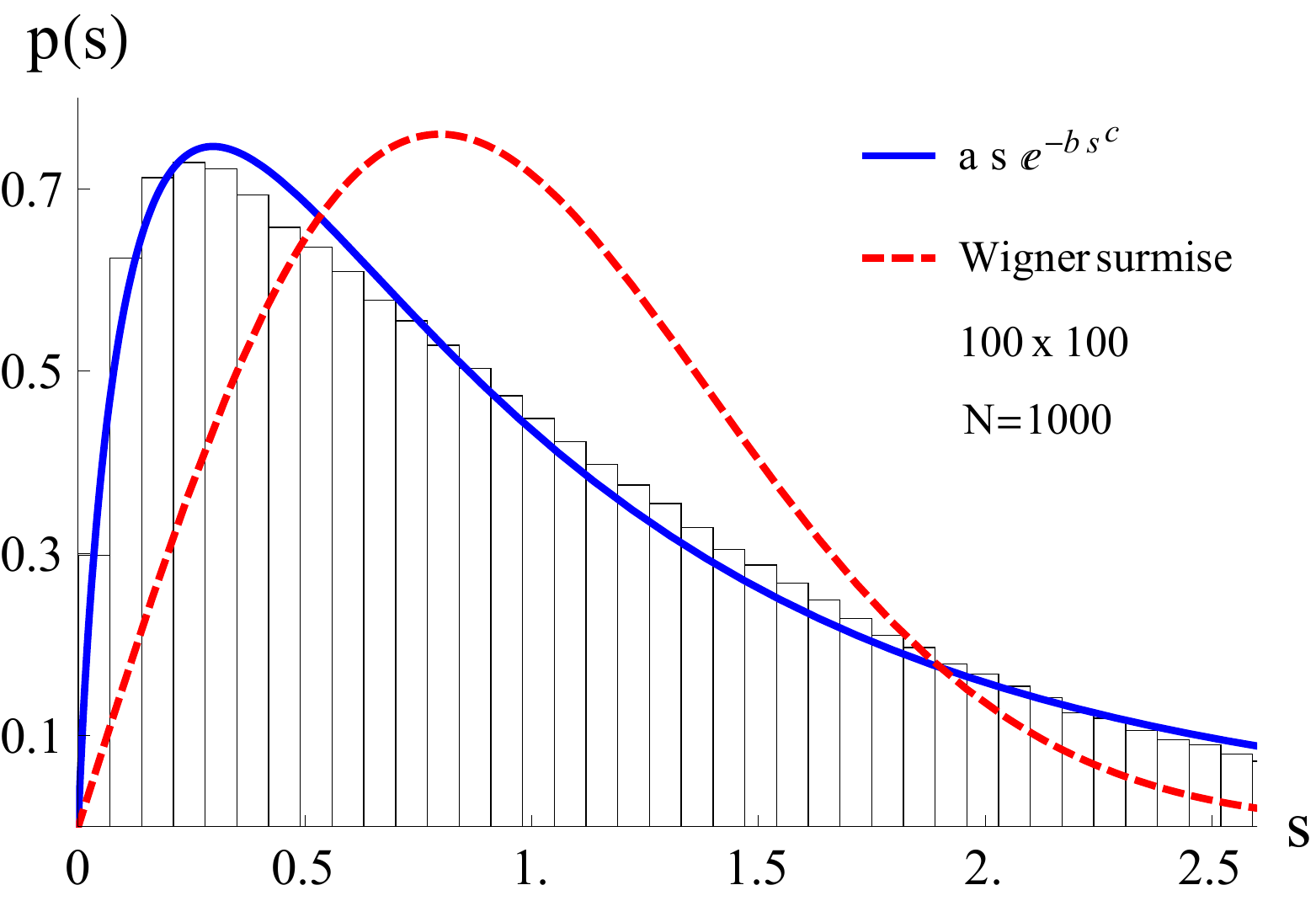}
	\includegraphics[width=7 cm,height=5.cm]{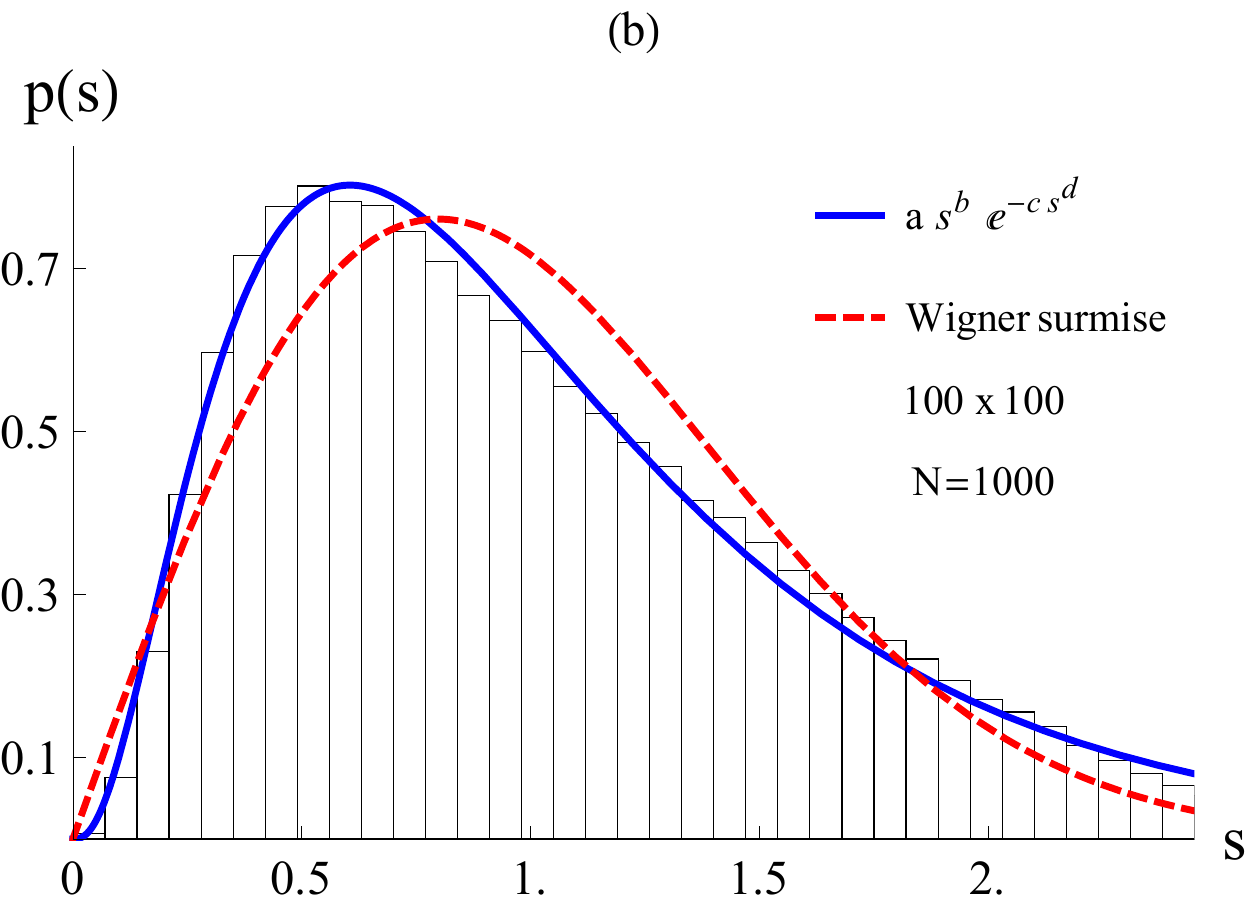}
	\caption{$p(s)$ for complex eigenvalues in upper plane for non-symmetric cyclic matrices $C$ under Gaussian PDF. This result is almost insensitive to an increase in the number or oder of the matrices and the PDF. Here the parameters of the empirical fit are $a=14.4, b=3.50, c=0.57$. We get almost similar result for non-symmetric tridiagonal matrix $T$, where $a=4.42,b=2.31,c=0.89$. The dotted curve represents Wigner's surmise. (b): The same for non-symmetric matrix real matrix $R$. Note that matrix $R$ has $n^2$ real random elements, hence a different behaviour of $p(s)$ than in (a). The values of parameters of the empirical fit in this case are $a=895, b=3.10, c=7.26, d=0.57.$ Notice $\sim s^3$ behaviour near $s=0$.}  
	\end{figure} 

\begin{table}[]
	\centering
	\caption{The $\mu$ values for the best fits of histograms to $p_{\mu}(s)$ (27). For ${\cal R'}$, ${\cal R}$ we use NLE (Nearest Levels of Ensemble); for  ${\cal C}, {\cal Q}$, ${\cal T}$, $T'$
		and $\Theta$ we use NLM (Nearest Levels of Matrix). }
	\label{my-label}
	\begin{tabular}{|c|c|c|c|c|c|c|c|}
		\hline
		$\hspace{0.05cm}$ $f(x)$ $\hspace{0.05cm}$ &   $\hspace{0.05cm}$  $\mathcal{R}$ $\hspace{0.05cm}$   & $\mathcal{R}$ & $\mathcal{Q}$ &  $\hspace{0.05cm}$  $\mathcal{C}$ $\hspace{0.05cm}$   & $\mathcal{T}$ $\hspace{0.05cm}$  &   $\hspace{0.05cm}$  $T'$ $\hspace{0.05cm}$ &   $\hspace{0.05cm}$  $\Theta$ $\hspace{0.05cm}$ \\ \hline

		G    & $\hspace{0.05cm}$ 1.14 $\hspace{0.05cm}$  & $\hspace{0.05cm}$ 1.14 $\hspace{0.05cm}$                          & $\hspace{0.05cm}$ 1.19 $\hspace{0.05cm}$                           &$\hspace{0.05cm}$  1.29    $\hspace{0.05cm}$   &$\hspace{0.05cm}$  1.12    $\hspace{0.05cm}$ &$\hspace{0.05cm}$  1.20    $\hspace{0.05cm}$    &$\hspace{0.05cm}$  1.35    $\hspace{0.05cm}$                        \\ \hline
		U                   &1.14  & 1.15                           & 1.17                           & 1.27                           & 0.99    &1.02   & 1.34   \\ \hline
		E                   &1.17  & 1.16                           & 1.24                           & 1.28                           & 1.11     &1.09   &1.32  \\ \hline
		T                  &1.14   & 1.14                           & 1.18                           & 1.26                           & 1.05     & 1.25   & 1.35 \\ \hline
		P                   & 1.15  & 1.15                           & 1.18                           & 1.29                           & 1.04    & 1.13   & 1.34  \\ \hline
		S                    & 1.13 & 1.15                           & 1.17                           & 1.28                           & 1.01    & 1.07   & 1.35  \\ \hline
		SG                    & 1.14 & 1.15                           & 1.18                           & 1.27                           & 1.07    & 1.06   &1.35   \\ \hline
		
	\end{tabular}
\end{table}

	\begin{table}[]
		\centering
		\caption{The $\mu$ values for the best fits of histograms to $p_{\mu}(s)$ (27). For non-symmetric matrices $R$, $C$, and $T$ we use $\Re(E_n), \Im(E_n)$ and $|E_n|$ in NLM. For various PDFs, G: Gaussian, U: Uniform, E: Exponential, SG: Super-Gaussian, T: Triangular, P: Parabolic; see the text at the end of Introduction.}
		\label{my-label}
		\scalebox{0.83}{
			\begin{tabular}{|c|c|c|c|c|c|c|c|c|c|}
				\hline
				$\hspace{0.005cm}$ \multirow{2}{*}{$f(x)$} $\hspace{0.005cm}$ & \multicolumn{3}{c|}{$R$}     & \multicolumn{3}{c|}{$C$}  & \multicolumn{3}{c|}{$T$} \\ \cline{2-10} 
				& $\hspace{0.005cm}$ Re{[$\bullet$}{]} $\hspace{0.005cm}$ & $\hspace{0.005cm}$  Im{[$\bullet$}{]} $\hspace{0.005cm}$ & $\hspace{0.005cm}$ $|\bullet|$ $\hspace{0.005cm}$   & $\hspace{0.005cm}$ Re{[$\bullet$}{]} $\hspace{0.005cm}$ &$\hspace{0.005cm}$  Im{[$\bullet$}{]} $\hspace{0.005cm}$ & $\hspace{0.005cm}$ $|\bullet|$ $\hspace{0.005cm}$ & $\hspace{0.005cm}$ Re{[$\bullet$}{]} $\hspace{0.005cm}$ &$\hspace{0.005cm}$  Im{[$\bullet$}{]} $\hspace{0.005cm}$ & $\hspace{0.005cm}$ $|\bullet|$ $\hspace{0.005cm}$   \\ \hline
				G                                 & 0.49     & 0.89     & 0.59 & 0.65     & 1.35     & 0.63 & 0.92     & 0.60     & 0.96 \\ \hline
				U                                 & 0.49     & 0.89     & 0.59 & 0.65     & 1.35     & 0.63 & 0.79     & 0.58     & 0.85 \\ \hline
				E                                   & 0.49     & 0.89     & 0.58 & 0.65     & 1.35     & 0.63 & 0.83     & 0.62     & 0.87 \\ \hline
				T                                  & 0.49     & 0.90     & 0.59 & 0.63     & 1.35     & 0.63 & 0.85     & 0.58     & 0.87 \\ \hline
				P                                   & 0.49     & 0.90     & 0.59 & 0.64     & 1.34     & 0.63 & 0.84     & 0.58     & 0.88 \\ \hline
				S                                   & 0.49     & 0.89     & 0.59 & 0.65     & 1.35     & 0.62 & 0.82     & 0.59     & 0.86 \\ \hline
				SG                                 & 0.49     & 0.89     & 0.59 & 0.64     & 1.35     & 0.62 & 0.83     & 0.59     & 0.87   \\ \hline
			\end{tabular}}
		\end{table}

Lastly, we plot the histograms of distribution of eigenvalues $D(\epsilon)$ where $\epsilon=E/E_{max}$ by taking $10^4 \times 10^4$ random real matrices which have real eigenvalues; all of these are symmetric excepting $T'$ which is pseudo-symmetric. For full fledged symmetric matrices ${\cal R}$ and ${\cal R}'$ which have $n(n-1)/2$ iid matrix elements we recover Wigner's well known semi-circle law. Next for symmetric cyclic matrix ${\cal C}$, the fit to the histogram by the function
\begin{equation}
 D(\epsilon)=a(1-e^{-a})^{-1}~|\epsilon|~ e^{-a\epsilon^2},~ a>0, ~ -1 \le \epsilon \le 1,
 \end{equation} 
 $\epsilon=E/E_{max}$. It is  
  consistent with the proposed limiting form (5)  [17]. Our calculations show $D(0) \ne 0$ for $n=10^4$, however we check that $D(0)\rightarrow 0$ when $n$ is far to large. The convergence of $D(0)$ to 0 is extremely slow. The parameter $a$ depends up on the order of the matrix in our case in Fig. 11(b) we get $a=7.08$. For Toeplitz random matrix $\Theta$, we find
the $D(\epsilon)$ is fitted well by a Gaussian: $1.20 e^{-4.23 \epsilon^2}$, this we believe is a new universality class in the distribution of eigenvalues of random matrices (see Fig. 11(c)).
We find that the histograms of $D(\epsilon)$ for the symmetric ${\cal T}$ and the pseudo symmetric ($T'$)  tridiagonal random matrices are fitted well by the empirical super-Gaussian: $0.94 e^{-8.06 \epsilon^4}$  and $1.03 e^{-11.63 \epsilon^4}$ (see Fig. 11(d,e)). Real matrices of the types $MM^t$ and $M^tM$ are positive-definite having positive eigenvalues provided their determinant is non-zero [21]. The secondary random symmetric matrices ${\cal Q}, {\cal D}$ and ${\cal S}$ are positive definite  having all  eigenvalues as positive commonly show a new distribution which is exponential: $D(\epsilon)=9.33~ e^{-9.33\epsilon}$ (see Fig. 11(f)).
For ${\cal Q}$ and ${\cal S}$ we get $D(\epsilon)$ as sub-exponential.
\begin{figure}[H]
	\centering
	\includegraphics[width=2.8cm,height=3 cm]{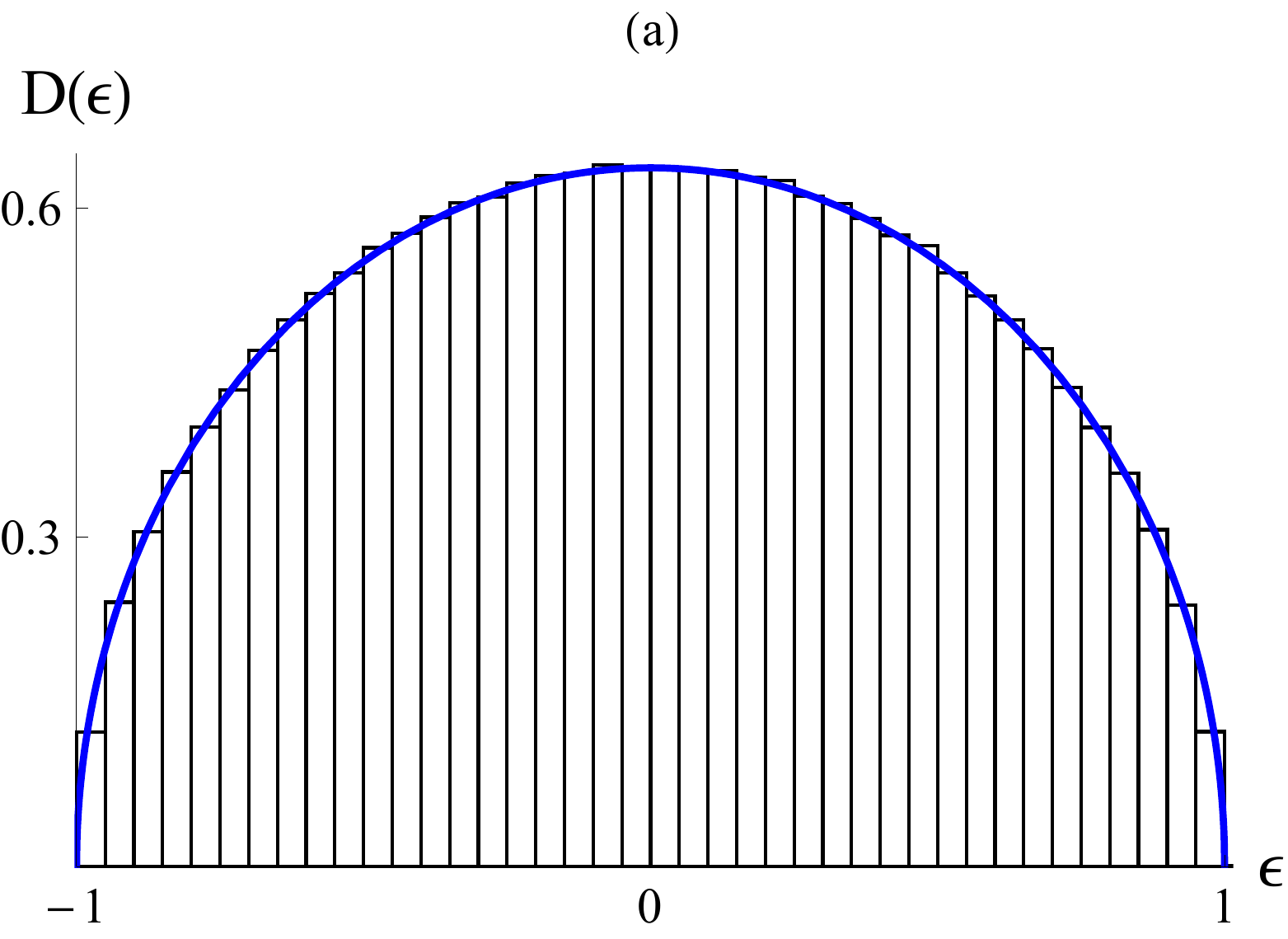}
	\includegraphics[width=2.8cm,height=3.cm]{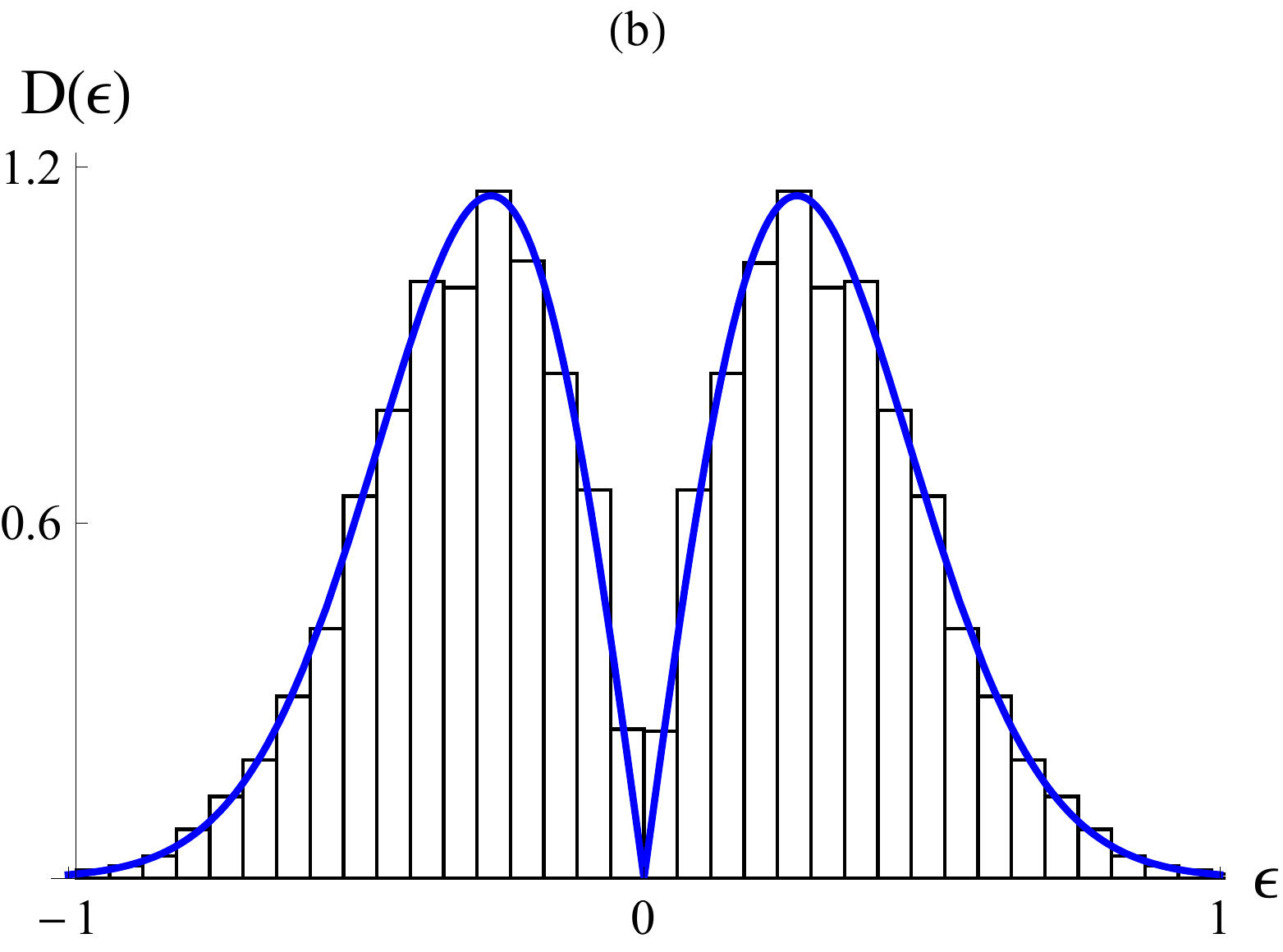}
		\includegraphics[width=2.8 cm,height=3.cm]{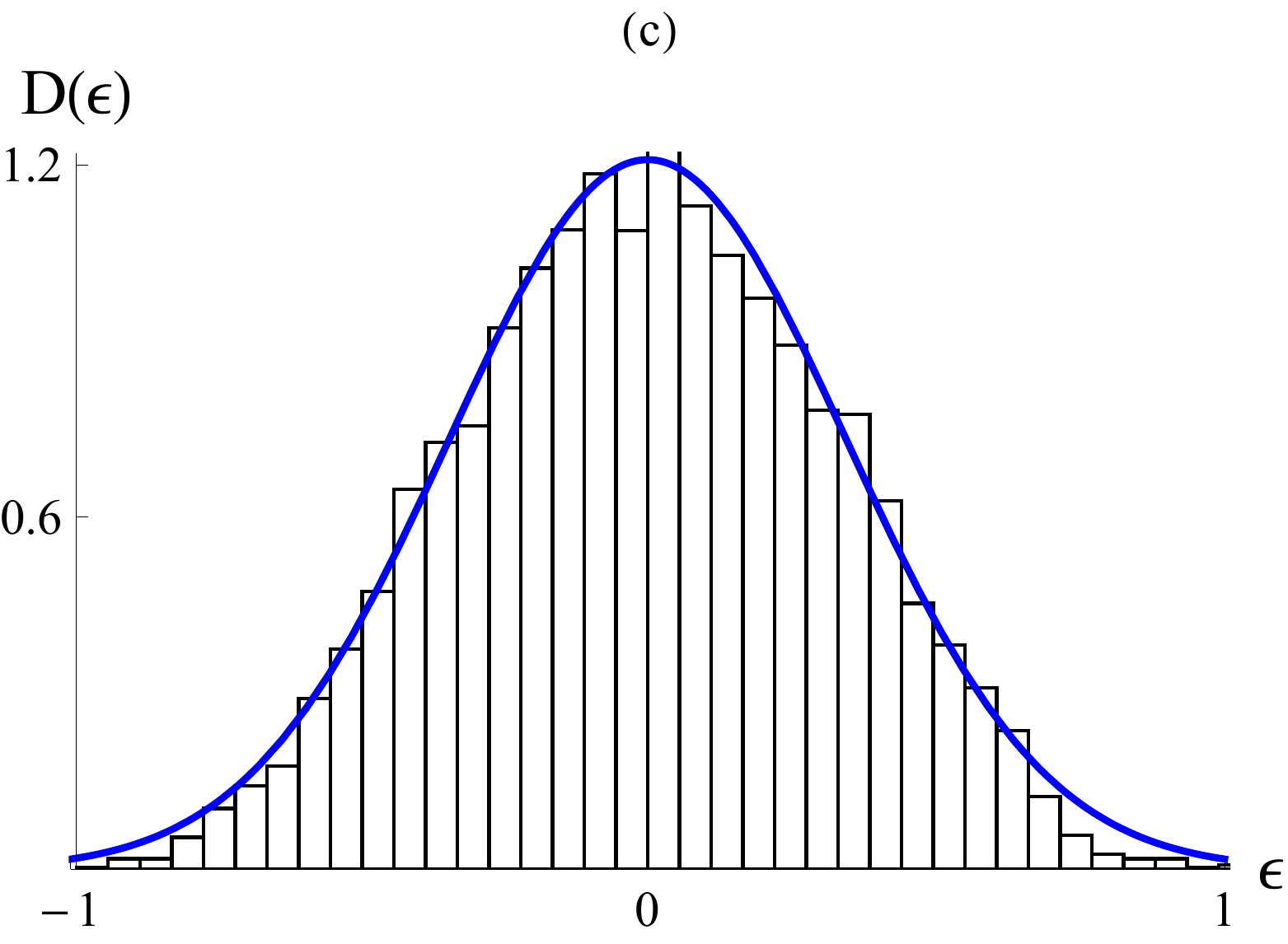} \\
%\end{figure} 
%\begin{figure}[]
	\centering
	\includegraphics[width=2.8cm,height=3. cm]{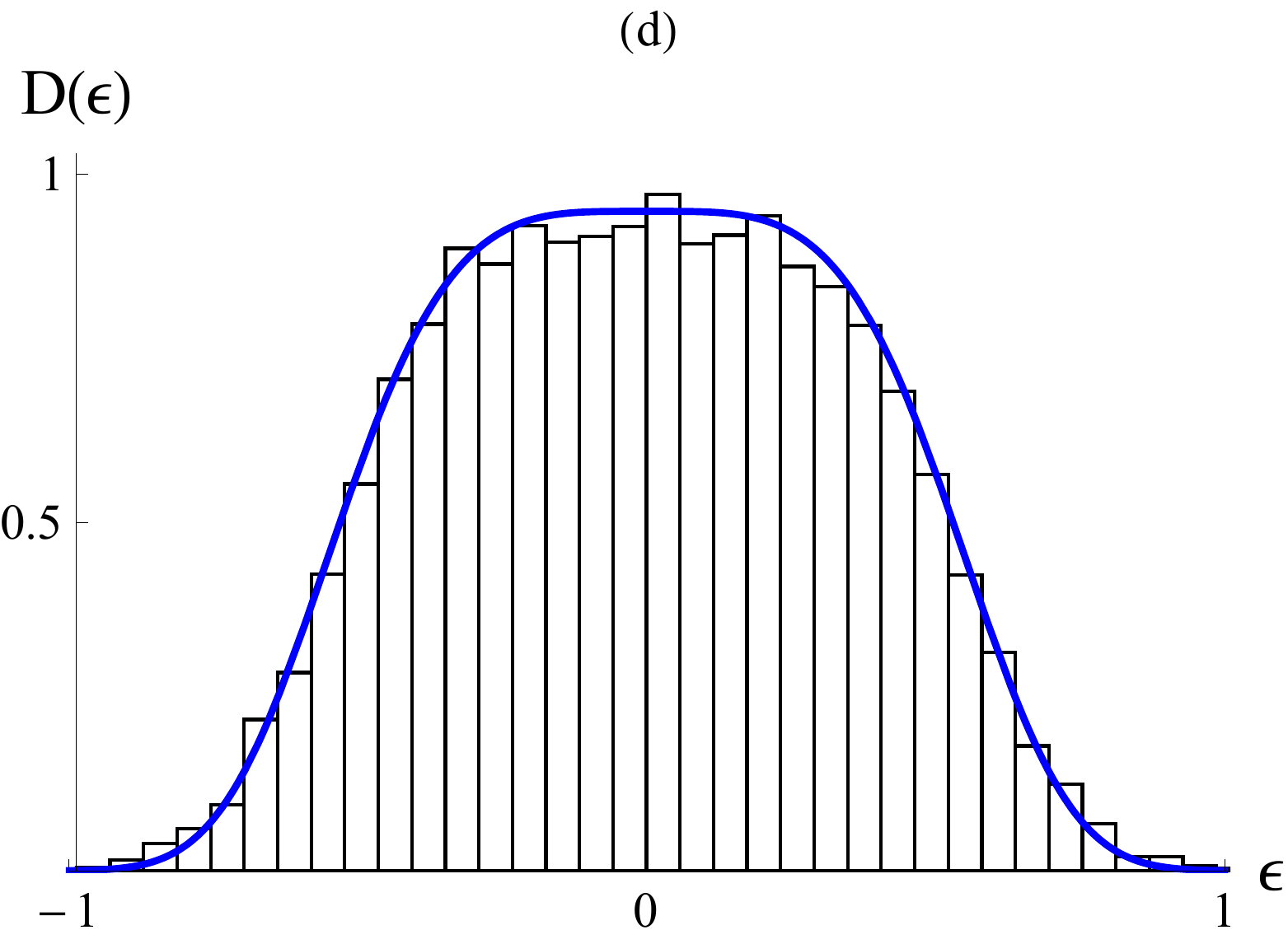}
	\includegraphics[width=2.8cm,height=3.cm]{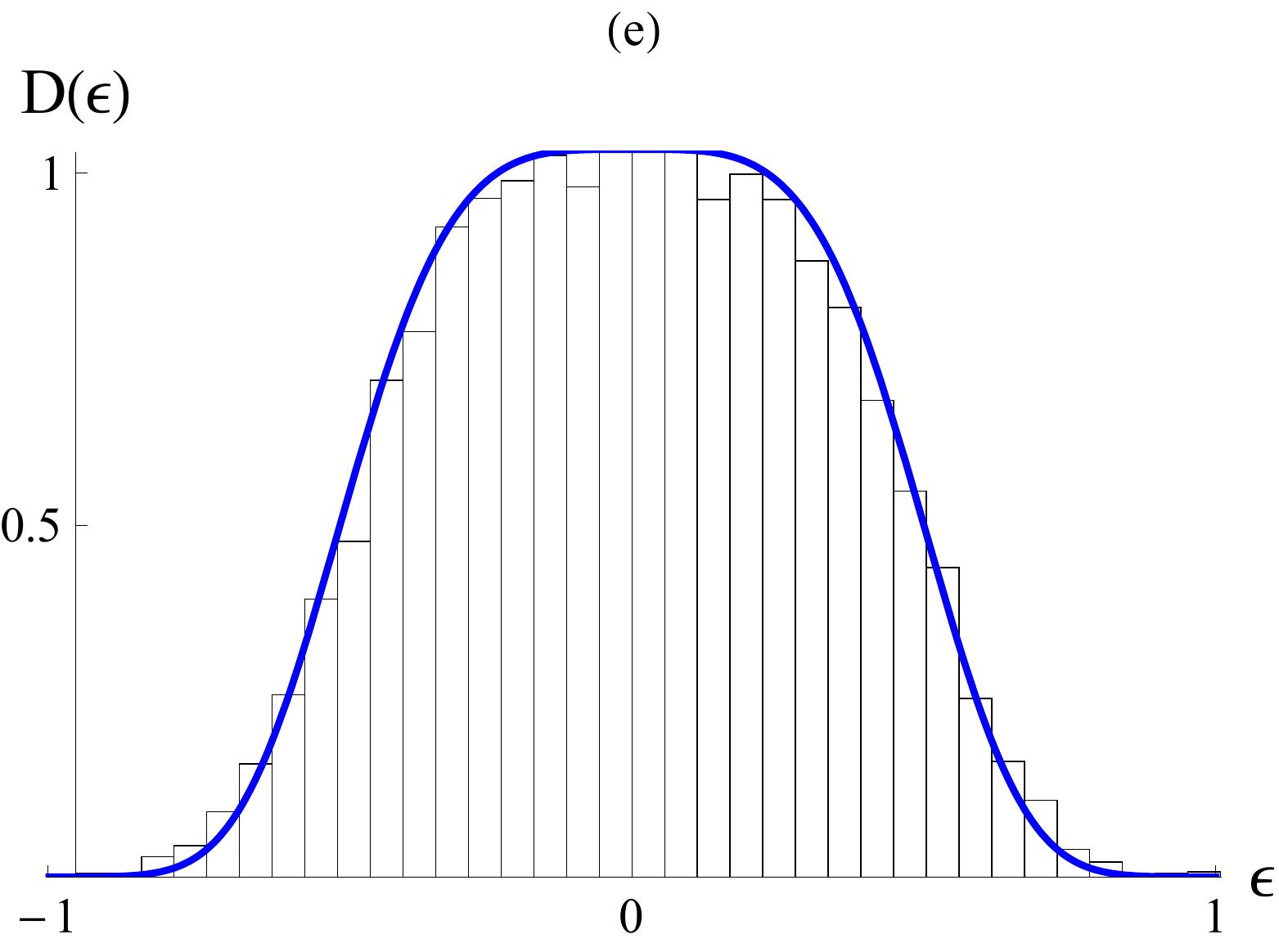}
	\includegraphics[width=2.8cm,height=3.cm]{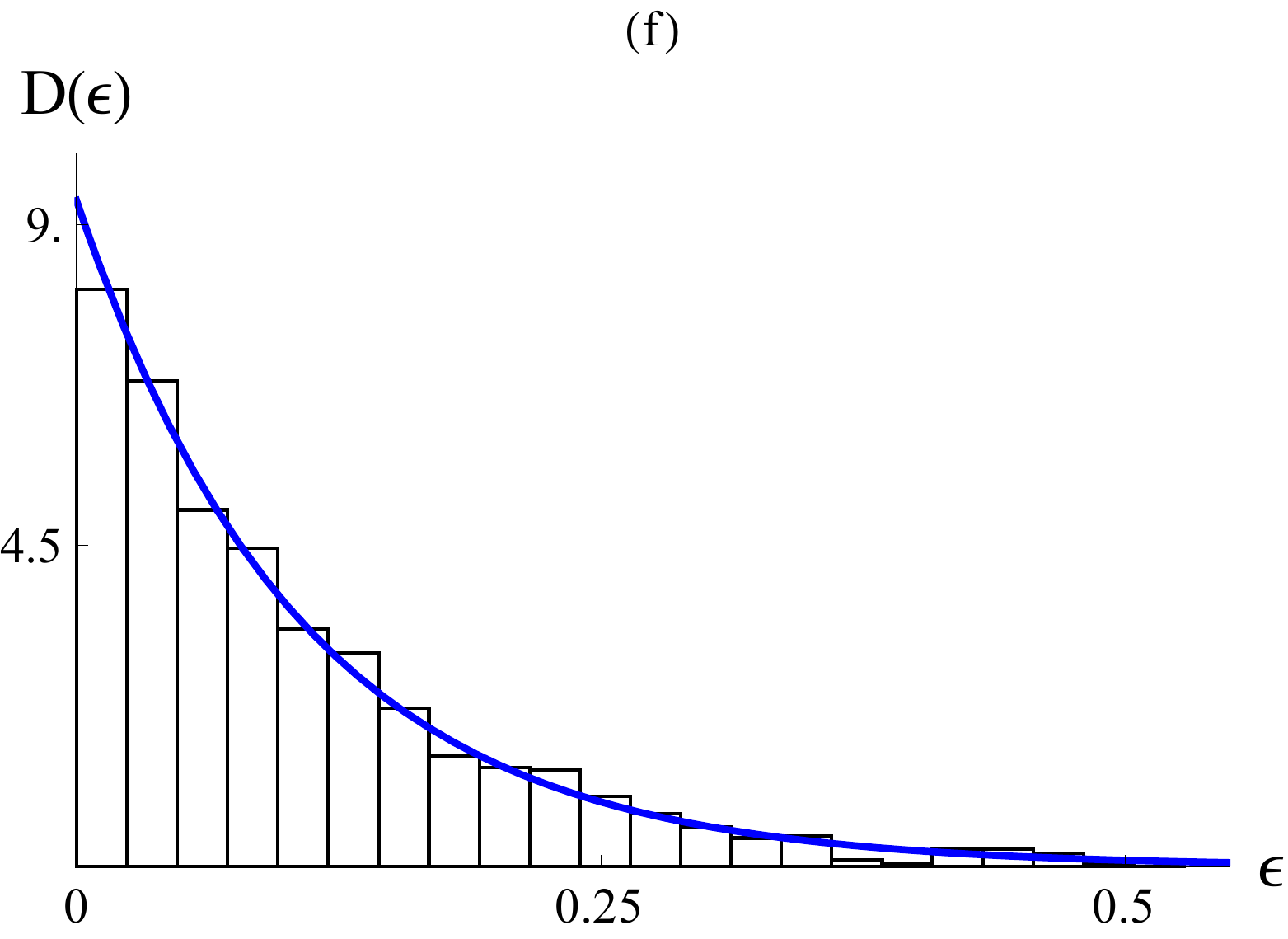}
\caption{The distribution of eigenvalues of $10^4\times 10^4$ Gaussian random real symmetric matrices discussed presently. (a): ${\cal R}, {\cal R}'$; (b): ${\cal C}$, (c) $\Theta$, (d): ${\cal T}$, (e):$T'$ and (f) represents $D(\epsilon)= 9.33 e^{-9.33 \epsilon}$ for the secondary symmetric matrix:${\cal D}$. Here, we have deliberately used the number of replicas of these matrices as one ($N=1)$. In fact for $N>>1$, our empirical fits (solid blue lines) look even better.}
\end{figure}

\section{V. Conclusions:}

Our analytic and semi-analytic results on $P(S)$ for two modifications of $2 \times 2$ real symmetric matrices in Eqs. (8,11,14,16,19,20) under various probability distribution functions
and the plotted $p(s)$ in Figs. (1-3) are new and instructive. They all give the linear level repulsion as $\alpha s$ near $s=0$ but notably $\alpha$ is not fixed. The distribution of eigenvalues for two real matrices under Gaussian PDF obtained in Eqs. (24,25) are also new and instructive.

Our present paper can be seen as an update on  the ensembles of various real matrices where Wigner's surmises for both spectral statistics $p(s)$ and $D(\epsilon)$ (3,4) have been encountered only when the real matrix is symmetric with $n(n-1)/2$ iid matrix elements and the probability distribution of elements is symmetric. The dramatic effect of making the PDF asymmetric is displayed in Fig. 5. For the symmetric PDF $P$ we get $p_{AB}(s)$ where  $A/2 \approx B \approx \pi/4$. However, for asymmetric cases $P_2$ and $P_3$ we observe $A/2 \approx B >>\pi/4$ (also see Table II for $P_2, P_3$ and Table I for $P$). In Table II,  see larger values of $A, B$ for ${\cal R}'$ when half-Gaussian, half-Uniform,...distributions are used. Also notice the sensitivity of $A,B$ on the choice of PDF.

The occurrence of Poisson statistics (see Table III) for the spacing of mixed levels of ensemble (NLE) of ${\cal R}$ and ${\cal R}'$ is expected but the occurrence of the same for the levels of matrix (NLM) for the ensemble of randm matrices: symmetric cyclic (${\cal C}$), symmetric tridiagonal ${\cal T}$), symmetric Toeplitz ($\Theta$) and pseudo-symmetric tridiagonal $(T')$ is a new result which we attribute to much less than $n^2/2$ number iid elements in these matrices. Interestingly, these random matrices give rise to different distribution of eigenvalues $D(\epsilon)$ in Fig. 11, where there are four new results Fig. 11(c-f).

The secondary symmetric matrices ${\cal Q}=RR^t, {\cal D}= CC^t , {\cal S}=TT^t$  have  positive definite eigenvalues, their distribution of eigenvalues $D(\epsilon)$ is new and it is  of exponential type for ${\cal D}$ (see Fig. 11(f)) and for ${\cal Q}$ and ${\cal S}$ it turns out to be sub-exponential. We find that $p(s)$ for ${\cal D}$ and ${\cal T}$ is sub-exponential
(Fig. 8) as their matrix elements are correlated and their elementary partners $C$ and $T$ have much less than $n^2/2$ iid elements. In this regard ${\cal Q}$ behaves differently by obeying exponential statistics for $p(s)$, as $R$ has $n^2(>n^2/2$) iid elements.

Our observation of Poisson statistics $p_{\mu}(s)$ when we use real, imaginary and modulus of complex eigenvalues $E^c_k$ in NLM for non-symmetric real matrices $R$  $C$ and $T$ is amusing. Interestingly, the $\mu$ values show a consistent trend and they are almost insensitive to the choice of PDF.

Our attempt to associate nearest level spacing statistics to  non-symmetric real matrices $C, T, R$
with complex eigenvalues in full and half plane give rise to new spacing distributions. Remarkably, $C$ and $T$ show similar results in Figs. (9(a), 10(a)) due to ($<n^2/2$) $n$ and $3n-2$ number of matrix elements which are iid. On the other hand, the matrix $R$ with $n^2$ number of iid matrix elements presents different scenarios in Figs. (9(b),10(b)). These $p(s)$ are remarkably different from Wigner's surmise and they are also insensitive to the choice of PDFs.

Pseudo-symmetric matrices represent Parity-Time-symmetric Hamiltonians[10-12,22] which may have some (others as complex-conjugate pairs) or all eigenvalue as real. In the former case PT-symmetry is broken. In the latter it is exact so these Hamiltonians may act as real-symmetric. It is due to this reason the spectral statistics of ${\cal T}$ (symmetric) and $T'$(pseudo-symmetric) are similar (exponential). But the ensemble of pseudo-symmetric matrices $C$ (with only two real eigenvalues per matrix) presents most different (half-Gaussian) level spacing distribution $p(s)$, where in the maximum occurs at $s=0$. However, its $D(\epsilon)$ turns out to be (Gaussian)-- similar to the ensemble of Toeplitz matrices and remarkably different to those of $2\times 2$ real matrices (1) presented in Eqs. (24,25) and Fig. 4. The results on the spectral statistics  of $N$, $n\times n$ ($n,N$ large) pseudo-symmetric/pseudo-Hermitian are eagerly awaited, in this regard, the half-Gaussian $p(s)$ for the ensemble of cyclic matrices (with two real eigenvalues) presented in Fig. 7(a), could be seen
as the first result.

Finally, we hope that numerical results on the spectral statistics: $p(s)$ and $D(\epsilon)$ for  ensembles of  large number of replicas of various random real matrices of large order presented here would inspire important analytic results in  future.

\end{document}